\documentclass[%
 reprint,
 amsmath,amssymb,
 aps,
]{revtex4-1}
\usepackage{natbib}
\usepackage{placeins}
\usepackage{chngcntr}
\usepackage{cleveref}
\usepackage{amsmath}
\usepackage{fancyhdr}
\usepackage{graphicx}% Include figure files
\graphicspath{C:/Users/Arpan/Desktop/Graphene_paper/Paper_figures}
\usepackage{dcolumn}% Align table columns on decimal point
\usepackage{bm}% bold math
\usepackage{lipsum}
\def\bibsection{\section*{References}}

\newcommand{\RN}[1]{%
  \textup{\uppercase\expandafter{\romannumeral#1}}%
}
\begin{document}
\title{Magnetothermoelectric effects in graphene and their dependence on scatterer concentration, magnetic field and band gap}

\author{Arpan Kundu}
\author{Majed Alrefae}
\author{Timothy S. Fisher}
\begin{abstract}
\section*{Abstract}
Using a semiclassical Boltzmann transport equation (BTE) approach, we derive analytical expressions for  electric and thermoelectric transport coefficients of graphene in the presence and absence of a magnetic field.  Scattering due to acoustic phonons, charged impurities and vacancies are considered in the model. Seebeck ($S_{xx}$) and Nernst ($N$) coefficients have been evaluated as functions of carrier density, temperature, scatterer concentration, magnetic field and induced band gap, and the results are compared with experimental data. $S_{xx}$ is an odd function of Fermi energy while $N$ is an even function, as observed in experiments. The peaks of both coefficients are found to increase with decreasing scatterer concentration and increasing temperature. Furthermore, opening a band gap decreases $N$ but increases $S_{xx}$. Applying a magnetic field introduces an asymmetry in the variation of $S_{xx}$ with Fermi energy across the Dirac point. The formalism is more accurate and computationally efficient than the conventional Green's function approach used to model transport coefficients and can be used to explore transport properties of other exotic materials.
\end{abstract}

\pacs{Valid PACS appear here}% PACS, the Physics and Astronomy
\maketitle

%\tableofcontents
\section{\label{sec:level1}Introduction}
Thermoelectric materials are capable of continuous generation of electric voltage in response to a temperature gradient without involvement of moving parts. Many prior research efforts have focused on developing new materials for practical power generation applications based on thermoelectrics since the discovery of Bi$_2$Te$_3$ in the 1950s \cite{Goldsmid_1954}. In comparison to thermoelectric effects, thermomagnetic effects have been less studied. The Nernst (Ettingshausen) effect is the phenomenon of generation of electric (heat) current in the cross product direction between a temperature gradient (electric field) and an applied magnetic field. The advantages of building a thermomagnetic conversion module have been highlighted by Sakuraba et al.\ \cite{Sakuraba2013}. Their module consisting of series-connected ferromagnetic wires has a much simpler design than the conventional thermoelectric module consisting of alternatively aligned n- and p- type semiconductor pillars. Apart from devices, these effects have also been used as sensitive probes for studying the electronic structure of various materials \cite{Hwang2009}. \par 
Graphene is a 2D material with high electron mobility and mechanical strength \cite{novoselov2005}. These attributes make it a good candidate for thermoelectric and thermomagnetic applications. However, it has two major disadvantages in this regard. First, it is gapless and has a symmetric electronic bandstructure, and consequently the Seebeck coefficient $S_{xx}$ is zero near the Dirac point corresponding to opposite contributions from electrons and holes. Secondly, it is an excellent thermal conductor, resulting in a low value of $\textit{ZT}$ \cite{2015_Dolfus}. \par
However, several studies \cite{2011_Xiao,2011_Wang_2} have demonstrated that appropriate nanostructuring and introduction of a band gap around the Dirac point can enhance the in-plane Seebeck coefficient  many-fold without degrading electrical conductivity. Xiao et al.\ \cite{2011_Xiao} demonstrated an enhancement of Seebeck coefficient in few-layer graphene (FLG) films up to 700 $\mu$V/K upon oxygen plasma treatment, which was attributed to disorder-induced band gap opening. Furthermore Wang et al.\ \cite{2011_Wang_2} obtained a Seebeck coefficient of ~180 $\mu$V/K upon introducing a band gap by applying a vertical electric field between two gates in dual-gated bilayer graphene. \par
 Experimentally, thermoelectric and thermomagnetic measurements are performed as functions of Fermi energy $E_F$. The Fermi energy is tuned by varying the gate voltage and thereby controlling the charge carrier density. Depending on whether $E_F$ is above or below the Dirac point, the system behaves either electron-like or hole-like. Most experimental studies have used single-layer graphene sheets mechanically exfoliated from highly oriented pyrolytic graphite (HOPG) \cite{Zuev2009,Wei2009,Checkelsky2009,Liu2012}. There are however a few reports for graphene sheets grown using chemical vapor deposition (CVD) \cite{Liu2012,Wang2011,2011_Xiao,2013_Babichev,Nam2014}  that are then transferred onto Si/SiO$_2$ substrates. Although mechanical exfoliation provides high-quality single-layer graphene, it cannot be used for mass production. On the other hand CVD growth is considered an attractive technique to mass produce graphene. \par
Table \RN{1}   lists the experimental parameters for Seebeck and Nernst effect measurements in the literature. In most cases, a microfabricated heater is used to create a temperature gradient in the longitudinal (in-plane) direction. The thermovoltage signal  generated in the longitudinal and transverse directions is detected using a four-point method in a Hall bar geometry. Checkelsky and Ong \cite{Checkelsky2009} studied the thermoelectric properties of exfoliated graphene on SiO$_2$ in the presence of strong magnetic field (5-14 T). Both Seebeck and Nernst coefficients were found to oscillate at high magnetic fields implying the formation of discrete Landau levels. An increase in magnetic field from 5 to 14 T led to a rise in the peak value of the transverse Seebeck coefficient $S_{xy}$ from 20 to 40 $\mu$V/K  and a fall in the number of oscillations. Liu et al.\ \cite{Liu2012} measured the longitudinal Seebeck coefficient $S_{xx}$  and the transverse Seebeck coefficient $S_{xy}$ for samples with different mobilities for magnetic fields up to 8 T. The peak in $S_{xy}$ increased almost linearly from 22 $\mu$V/K at 1 T to 160 $\mu$V/K at 8 T, while the peak in $S_{xx}$  rose from 70 to 100 $\mu$V/K upon increasing the magnetic field from 2 to 8 T. \par
Theoretical modeling of Seebeck and Nernst effects has been primarily performed using either the analytical Mott relation or the non-equilibrium Green's function  (NEGF) approach. The Mott relation \cite{Zuev2009,Wang2011,2010_Zhu}  predicts a linear rise in $S_{xx}$ with temperature. Zuev et al.\ \cite{Zuev2009} found experimental results to follow the Mott relation in the temperature range of 10-300 K. They attributed this observation to the dominance of impurity scattering over phonon scattering. The Mott relation has been thus found to hold true only when $\frac{\gamma}{k_bT}>>1$ \cite{Wei2009} where $\gamma$ is the impurity bandwidth proportional to the square root of impurity concentration $\gamma \propto \sqrt{n_{imp}}$ \cite{2007_Lofwander}. In order to avoid inaccuracies at high temperatures, many authors have used NEGF methods to model thermoelectric and magnetothermoelectric effects \cite{Xing2009,Ugarte2011,Hinz2014,Yan2010}. Ugarte et al.\ \cite{Ugarte2011} included effects of unitary and charged impurities and presented results for both low and high temperatures. Although the NEGF method is more rigorous than the Mott relation, it cannot treat scattering of charged impurities at high magnetic fields. \par
A thorough understanding of the energy dependence of various scattering mechanisms is crucial  for elucidating thermoelectric and magnetothermoelectric transport properties \cite{Liu2012}. Different scattering mechanisms have been identified and experimentally validated for both suspended and unsuspended graphene films. These  include long range Coulomb scattering due to impurities in the SiO$_{2}$ substrate and electron-phonon scattering. Stauber et al.\ \cite{Stauber2007} proposed an additional scattering mechanism that may originate from vacancies, cracks or defects in the sample. They showed that incorporating this additional mechanism into the Boltzmann transport equation dramatically improved agreement between theoretical and experimental values of mobility. \par
In this work, we investigate the thermomagnetic and thermoelectric properties of single-layer graphene. The Boltzmann formalism has proven useful in understanding transport in graphene \cite{2007_Peres} as well as its three-dimensional counterpart, Weyl semimetal \cite{sharma2015} and shows good agreement with other theoretical approaches such as the Kubo formalism \cite{Sinitsyn_2007}. The analytical solution to the BTE, which is valid for a 1D system within the relaxation time approximation (RTA) and near equilibrium conditions, is used to derive analytical expressions for the various transport coefficients. The mean relaxation times accounting for various scattering mechanisms have been used in the derivations. The Seebeck coefficient $S_{xx}$  and Nernst coefficient $N$ have been thoroughly characterized as functions of impurity concentrations, temperature, carrier concentration and magnetic field, and valuable insights have been drawn by comparing model results with previously published experimental data. Although this technique of evaluatng thermoelectric and thermomagnetic properties has been applied to single layer periodic graphene in this paper, it can be used to study other materials in the future, given their bandstructre and scattering rate information. \par
The paper is organized as follows. In Sec. II, detailed formulation of the transport coefficients from the BTE solution is provided. The solution methodology and final form of the transport coefficients are presented in Sec. III. Model results are presented in Sec. IV, including parametric variations of impurity density, temperature, band gap and magnetic field. \par
\section{\label{sec:level2}Theory}
The Nernst coefficient $N$ is defined as \cite{1980_Moore,2009_Choiniere}
\begin{equation} \label{eq:1}
N=\frac{E_y}{-B\triangledown_xT}|_{J_x=J_y=0}
\end{equation} 
where $E_y$ is the electric field in the $y$ direction, $B$ is the magnetic field in the $z$ direction, and $\triangledown_xT$ is the temperature gradient in the $x$ direction. $J_x$ and $J_y$ are electrical current densities in the $x$ and $y$ directions respectively. The transverse Seebeck coefficient $S_{xy}$ is obtained by multiplying the Nernst coefficient $N$ with the magnetic field $B$. The longitudinal Seebeck coefficient $S_{xx}$ is
\begin{equation} \label{eq:2}
S_{xx}=\frac{E_x}{-\triangledown_x T}|_{J_x=J_y=B=0}
\end{equation}
 In linear response theory, $J_x$ and $J_y$ are related to the electric field and temperature gradient as
\begin{equation} \label{eq:3}
\begin{split}
J_x=\sigma_{xx}E_x+ \sigma_{xy}E_y + \alpha_{xx}\triangledown_xT \\
J_y=\sigma_{yy}E_y+ \sigma_{yx}E_x + \alpha_{yx}\triangledown_xT
\end{split}
\end{equation}
where $\sigma_{xx}$ and $\sigma_{xy}$ are the longitudinal and transverse electrical conductivities respectively, and $\alpha_{xx}$ and $\alpha_{xy}$ represent the longitudinal and transverse thermoelectric coefficients respectively. Under isotropic conditions $\sigma_{xx}=\sigma_{yy}$, $\alpha_{xx}=\alpha_{yy}$, $\sigma_{yx}=-\sigma_{xy}$ and $\alpha_{yx}=-\alpha_{xy}$. Setting the current densities to zero and using Eqs. (1) and (2), we obtain the expressions for Nernst and Seebeck coefficients as
\begin{equation} \label{eq4}
N=\frac{\sigma_{xy}\alpha_{xx}-\sigma_{xx}\alpha_{xy}}{B(\sigma_{xx}^2+\sigma_{xy}^2)}
\end{equation}
and 
\begin{equation} \label{eq5}
S_{xx}=\frac{\sigma_{xx}\alpha_{xx}+\sigma_{xy}\alpha_{xy}}{\sigma_{xx}^2+\alpha_{xy}^2}.
\end{equation}
The electronic states in periodic graphene at the six corners $K$ and $K'$ of the first Brillouin zone have a linear dispersion relation $\epsilon (k)=\pm \hbar v_F|k|$=$\pm \hbar v_F\sqrt{k_x^2+k_y^2}$, where $v_F$ is the Fermi velocity (assumed to be 1$\times$10$^6$ m/s \cite{2008_Bolotin, Geim_2007}). The density of states is
\begin{equation} \label{eq6}
D(\epsilon)=\frac{2|\epsilon|}{\pi \hbar^{2}v_F^2}
\end{equation}
The semiclassical Boltzmann transport equation (BTE) with the relaxation time approximation is
\begin{equation} \label{eq7}
\frac{\partial f_{\bf{k}}}{\partial t}+\vec{v}_{\bf{k}}.\vec{\triangledown}_{\bf{r}}f_{\bf{k}}+\vec{F_{e}}.\vec{\triangledown}_{\bf{p}}f_{\bf{k}}=-\frac{f_{\bf{k}}-f_{eq}}{\tau_{m}}
\end{equation}
\\
where $f_{\bf{k}}$ is the probability density function of carriers with wave vector $\vec{\bf{k}}$,  $f_{eq}$ is the equilibrium Fermi-Dirac distribution, $\tau_{m}$ is the mean relaxation time,  $v_{\bf{k}}$ is the group velocity $v_{\bf{k}}$=$\hbar^{-1} \partial \epsilon$/$\partial \bf{k}$, and $\vec{F}_e$ is the force field. In the presence of electric and magnetic fields, the force field is
\begin{equation} \label{eq8}
\vec{F}_e=e\vec{E}+e\vec{v}\times \vec{B}
\end{equation}
The 1D steady state linearized solution of the BTE  in the absence of magnetic field ($B=0$) is  \cite{Ashcroft}
\begin{equation} \label{eq9}
f_{\textbf{k}}=f_{eq}+\tau_m(-\frac{\partial f_{eq}}{\partial \epsilon})v_{\bf{k}}.(-e\textbf{E}+\frac{\epsilon - \mu}{T}(-\bf{\triangledown T}))
\end{equation}
\\
at near-equilibrium conditions. The resulting current density ($J_x$) is
\begin{equation} \label{eq10}
\begin{aligned}
J_x = 2\sum_{p} \sum_{k}ev_x.f_{\bf{k}}
\end{aligned}
\end{equation}
where the factor of 2 represents spin degeneracy and $p$ represents summation over all branches. From  Eqs. (3), (9) and (10) we obtain
\begin{equation}  \label{eq11}
\sigma_{xx}=4e^2 \int {\textbf{[dk]}}v_x^2\tau_m(-\frac{\partial f_{eq}}{\partial \epsilon})
\end{equation}
\begin{equation}  \label{eq12}
\alpha_{xx}=-\frac{4e}{T}\int {\textbf{[dk]}}v_x^2\tau_m (\epsilon - \mu)(-\frac{\partial f_{eq}}{\partial \epsilon})
\end{equation}
where ${\textbf{[dk]}}=\frac{d^2k}{(2\pi)^2}=\frac{kdk}{2\pi}$ .\\
We now discuss the case of a perpendicular magnetic field $B \hat{z}$ in the presence of a londitudinal thermal gradient $\triangledown_{x} T \hat{x}$. The Boltzmann equation is modified in the following form to account for the influence of the magnetic field \cite{sharma2015}
\begin{equation}  \label{eq13}
f_{\textbf{k}}=f_{eq}+(-\frac{\partial f_{eq}}{\partial \epsilon})(\tau_{m}v_{x}\frac{\epsilon - \mu}{T}(-\triangledown_{x} T)+v_x\Lambda_x+v_y\Lambda_y)
\end{equation}
The correction factors $\Lambda_x$ and $\Lambda_y$ have been derived \cite{sharma2015} by satisfying the steady state form of Eq. (7) while using the force field from Eq. (8).
\begin{widetext}
\begin{equation}  \label{eq14}
\Lambda_x=eB\tau_{m}\triangledown_{x}T\frac{\epsilon-\mu}{T}\frac{[\frac{v_x}{m_{xy}}-\frac{v_y}{m}][-\frac{eBv_y}{m}+\frac{eBv_x}{m_{xy}}-\frac{v_x}{\tau_m}]+[\frac{v_x}{m}+\frac{v_y}{m_{xy}}][\frac{eBv_x}{m}-\frac{eBv_y}{m_{xy}}-\frac{v_y}{\tau_m}]}{[-\frac{eBv_y}{m}+\frac{eBv_x}{m_{xy}}-\frac{v_x}{\tau_m}]^2+[\frac{eBv_x}{m}-\frac{eBv_y}{m_{xy}}-\frac{v_y}{\tau_m}]^2} 
\end{equation}
\begin{equation}  \label{eq15}
\Lambda_y=eB\tau_{m}\triangledown_{x}T\frac{\epsilon-\mu}{T}\frac{[\frac{v_x}{m_{xy}}-\frac{v_y}{m}][-\frac{eBv_y}{m_{xy}}+\frac{eBv_x}{m}-\frac{v_y}{\tau_m}]-[\frac{v_x}{m}+\frac{v_y}{m_{xy}}][-\frac{eBv_y}{m}+\frac{eBv_x}{m_{xy}}-\frac{v_x}{\tau_m}]}{[-\frac{eBv_y}{m}+\frac{eBv_x}{m_{xy}}-\frac{v_x}{\tau_m}]^2+[\frac{eBv_x}{m}-\frac{eBv_y}{m_{xy}}-\frac{v_y}{\tau_m}]^2}
\end{equation}
\end{widetext}
Considering only linear terms in $B$
\begin{equation}
 \Lambda_i=\tau_m \triangledown_x T\frac{\epsilon-\mu}{T}c_i
\end{equation}
where $c_x=-\frac{eB\tau_m}{m_{xy}}$ and $c_y=\frac{eB\tau_m}{m}$.\@ The band mass $m_{ij}$ is defined as $m_{ij}^{-1}=\hbar^{-2} \partial^2 \epsilon(k)/\partial k_i \partial k_j$. Using Eqs. (3), (10) and (13), the thermoelectric coefficients (longitudinal and transverse) are
\begin{equation}  \label{eq16}
\alpha_{xx}=-\frac{4e}{T}\int \textbf{[dk]}v_x^2\tau_m (\epsilon - \mu)(-\frac{\partial f_{eq}}{\partial \epsilon})(1+\frac{eB\tau_m}{m_{xy}})
\end{equation}
\begin{equation}  \label{eq17}
\alpha_{xy}=-\frac{4e}{T}\int \textbf{[dk]}\tau_m(-\frac{\partial f_{eq}}{\partial \epsilon})eB\tau_m(\frac{v_y^2}{m}-\frac{v_xv_y}{m_{xy}})
\end{equation}
Similarly, the electrical conductivity coefficients (longitudinal and transverse) can be shown to be
\begin{equation}  \label{eq18}
\sigma_{xx}=4e^2\int \textbf{[dk]}v_x^2\tau_m(-\frac{\partial f_{eq}}{\partial \epsilon})(1+\frac{eB\tau_m}{m_{xy}})
\end{equation}
\begin{equation}  \label{eq19}
\sigma_{xy}= 4e^2\int \textbf{[dk]}\tau_m(-\frac{\partial f_{eq}}{\partial \epsilon})eB\tau_m(\frac{v_y^2}{m}-\frac{v_xv_y}{m_{xy}})
\end{equation}
under the isotropic approximation with $v_x=v_y=v_F/\sqrt{2}$ and $k_x=k_y=k_F/\sqrt{2}$. \par
We now discuss the various scattering mechanisms that have been considered in the analysis. Scattering due to long-range Coulomb scatterers in the Si/SiO$_2$ substrate exhibits a linear dependence with $k$ \cite{Stauber2007}
\begin{equation} \label{eq20}
\tau_{imp}(k)=\frac{16\hbar^2v_F\epsilon_0^2\epsilon_r^2(1+\gamma)^2|k|}{n_{imp}Z^2e^4}
\end{equation}
\\where $Z$ is the valency of the charged impurities considered to be 1, and $n_{imp}$ is the impurity concentration. Acoustic phonon scattering time varies inversely with $k$ \cite{Stauber2007}
\begin{equation} \label{eq21}
\tau_{AP}(k)=\frac{8\hbar^2 \rho v_s^2 v_F}{D_A^2k_bT|k|}
\end{equation}
$\tau_{AP,LA}$ and $\tau_{AP,TA}$  correspond to scattering with acoustic phonons of LA and TA branches respectively. $\rho$ is the mass per unit area (7.6$\times$ 10$^{-7}$ kg/m$^2$) and $v_s$ is the group velocity (7333 m/s and 2820 m/s) for the LA and TA phonon branches (respectively). $D_A$ is the acoustic deformation potential which generally varies with carrier concentration, but is assumed constant for simplicity in many studies. Bolotin et al.\ \cite{2008_Bolotin} found $D_A$ to be 29 eV for ultraclean suspended graphene compared to 17 eV reported for unsuspended graphene \cite{Chen2008}. On the other hand, Stauber et al.\ \cite{Stauber2007} assumed $D_A$ to be 9 eV. In this study, $D_A$ has been assumed to be 17 eV because unsuspended graphene is considered. Electron scattering with optical phonons are expected to have negligible contribution to the net scattering rate \cite{Stauber2007}. Acoustic phonon scattering and substrate impurity scattering have only been considered in the mean relaxation time for comparison with experimental findings in a few studies \cite{Adam2007,Tan2007}. However, Stauber et al.\ \cite{Stauber2007}  criticises this assumption because of the high density of charged impurities $n_{imp}>10^{12}$ cm$^{-2}$ needed in the Boltzmann formalism to obtain the experimentally observed mobilities, that are not likely to be present in an insulator like SiO$_2$. Based on this argument, they proposed an additional scattering mechanism due to midgap states that may be caused by vacancies present in the substrate. The scattering time of this mechanism is proportonal to $k$ up to logarithmic corrections
\begin{equation} \label{eq22}
\tau_{vac}(k)=\frac{|k|(ln(|k|R_0)^2}{\pi^2v_Fn_{vac}}
\end{equation}
where $n_{vac}$ is the vacancy concentration and $R_0$ is the average radius of vacancies assumed to be 1.4 {\AA}. On consideration of the foregoing three scattering mechanisms, a low impurity density of $n_{imp} \approx 10^{11}$ cm$^{-2}$ is able to justify the experimentally observed mobilities of $\mu \approx$ 5000 cm$^2$/(V.s). The mean relaxation time $\tau_m$ is finally obtained using Matthiesen's rule
\begin{equation} \label{eq23}
\frac{1}{\tau_{m}}=\frac{1}{\tau_{AP,LA}}+\frac{1}{\tau_{AP,TA}}+\frac{1}{\tau_{IMP}}+\frac{1}{\tau_{VAC}}
\end{equation}
Combining Eq.(21) to Eq.(24), $\tau_m$ is expressed as
\begin{equation} \label{eq24}
\frac{1}{\tau_m(k)}=\frac{\alpha_1}{|k|}+\alpha_2 |k|+\frac{\alpha_3}{|k|(ln(|k|R_0))^2}
\end{equation}
where $\alpha_1$, $\alpha_2$ and $\alpha_3$ are independent of $k$. $\tau_m$ is used in the Boltzmann formalism to obtain new expressions for the transport coefficients. To demonstrate the effect of an introduced bandgap, the following dispersion relation is assumed \cite{Patel2012}
\begin{equation} \label{eq25}
\epsilon(k)=\pm \sqrt{(\hbar^2v_F^2|k|^2+\bigtriangleup^2)}.
\end{equation}
where $\bigtriangleup$ is the bandgap. Such dispersion equation has been assumed to produce the bandstructure of Dirac systems possessing a band gap. In the above equation, $\epsilon$ has a minimum value of $\bigtriangleup$ at zero Fermi energy and it follows the dispersion relation of graphene at high $k$ values, considering $\bigtriangleup$ to be quite small. The group velocity for such a dispersion relation has an energy dependence
\begin{equation} \label{eq26}
v(\epsilon)=v_F\sqrt{1-\frac{\bigtriangleup^2}{\epsilon^2}}
\end{equation}
For a non-zero band gap, the transport coefficients are obtained by energy integration from -$\infty$ to +$\infty$ while omitting the range from -$\bigtriangleup$ to +$\bigtriangleup$, because the density of states is zero in that zone. Eq. (37) in the Appendix contains the final expression of the longitudinal conductivity $\sigma_{xx}$ for such a system.\par
\section{Solution Methodology}
Eqs. (28)-(31) shown below represent the final form of the transport coefficients used throughout rest of the paper. 
\begin{widetext}
\begin{equation}
\begin{split} \label{eq27}
\sigma_{xx}=&\frac{2e^2}{h}\frac{E_F^2|k_F|}{4\hbar k_bT}\int_0^\infty\frac{x^2(cosh^{-2}(\frac{E_F}{2k_bT}(x-1))+cosh^{-2}(\frac{E_F}{2k_bT}(x+1)))}{\alpha_1+\alpha_2k_F^2x^2+\frac{\alpha_3}{|ln(k_FxR_0)|^2}}dx \\
&-\frac{2e^2}{h}\frac{eBv_F^3k_F^3}{8k_bT}\int_0^\infty\frac{x^2(cosh^{-2}(\frac{E_F}{2k_bT}(x-1))-cosh^{-2}(\frac{E_F}{2k_bT}(x+1)))}{(\alpha_1+\alpha_2k_F^2x^2+\frac{\alpha_3}{|ln(k_FxR_0)|^2})^2}
\end{split}
\end{equation}
\begin{equation} \label{eq28}
\sigma_{xy}=\frac{2e^2}{h}\frac{eBv_F^3k_F^3}{4k_bT}\int_0^\infty\frac{x^2(cosh^{-2}(\frac{E_F}{2k_bT}(x-1))-cosh^{-2}(\frac{E_F}{2k_bT}(x+1)))}{(\alpha_1+\alpha_2k_F^2x^2+\frac{\alpha_3}{(ln(k_FxR_0))^2})^2}dx
\end{equation}
\begin{equation} 
\begin{split}  \label{eq29}
\alpha_{xx}=&\frac{2e^2}{h}\frac{v_F^2h|k_F|^3E_F}{8\pi ek_bT^2}[\int_0^\infty\frac{x^2(x-1)cosh^{-2}(\frac{E_F}{2k_bT}(x-1))-x^2(x+1)cosh^{-2}(\frac{E_F}{2k_bT}(x+1))}{(\alpha_1+\alpha_2k_F^2x^2+\frac{\alpha_3}{(ln(k_FxR_0))^2})}dx \\
&-\frac{2e^2}{h}\frac{Bv_F^3|k_F|^3E_F}{8k_bT^2}[\int_0^\infty\frac{x^2(x-1)cosh^{-2}(\frac{E_F}{2k_bT}(x-1))+x^2(x+1)cosh^{-2}(\frac{E_F}{2k_bT}(x+1))}{(\alpha_1+\alpha_2k_F^2x^2+\frac{\alpha_3}{(ln(k_FxR_0))^2})^2}dx
\end{split}
\end{equation}
\begin{equation} \label{eq30}
\alpha_{xy}=\frac{2e^2}{h}\frac{v_F^3BE_F|k_F|^4}{4k_bT^2k_F}\int_0^\infty\frac{x^2(x-1)cosh^{-2}(\frac{E_F}{2k_bT}(x-1))+x^2(x+1)cosh^{-2}(\frac{E_F}{2k_bT}(x+1))}{(\alpha_1+\alpha_2k_F^2x^2+\frac{\alpha_3}{(ln(k_FxR_0))^2})^2}dx
\end{equation}
\end{widetext}
The constants $\alpha_1$,$\alpha_2$ and $\alpha_3$ are given by
\begin{equation}
\begin{split}
\alpha_1 &=\frac{n_{imp}Z^2e^4}{16\hbar^2v_F\epsilon_0^2\epsilon_r^2(1+\gamma)^2} \\
\alpha_2 &=\frac{D_A^2k_bT(1/v_{s,LA}^2+1/v_{s,TA}^2)}{8\hbar^2 \rho v_F} \\
\alpha_3 &=\pi^2v_Fn_{vac} .
\end{split}
\end{equation}
 A sample derivation of $\sigma_{xx}$ is given in the Appendix. In these expressions, the integration variable $x$ represents the ratio $\epsilon/E_F$ where $E_F$ is the Fermi energy. The integrations are carried out using the trapezoidal rule. A step size of $10^{-4}$ is used for $x$ after mesh refinement analysis. The upper limit of $\infty$ in the integrations is taken as 10 to make them computationally efficient without affecting the results.\par Experimentally, the carrier concentration is controlled by varying the gate voltage ($V_g$) in a parallel plate geometry. For the experimental range of parameters used, it is generally assumed that the capacitance of the device is constant and the carrier density depends linearly on the gate voltage as $n=C_gV_g/e$, where the gate capacitance $C_g$ is the capacitance per unit area generally in the range of 100-115 aF/$\mu$m$^2$ \cite{Wei2009,Liu2012,Wang2011}. Here we assume $C_g$ to have a value of 100 aF/$\mu$m$^2$. The Fermi energy $E_F$ is related to the carrier density $n$ by $E_F=\hbar v_F\sqrt{\pi n}$.
\par
\section{\label{sec:level3}Results and Discussion}
We first consider the electrical conductivity coefficients at zero magnetic field. Figure 1(a) shows the temperature dependence of the different scattering rates. Figure 1(b) shows the temperature dependence of mobility $\mu$ calculated using the Drude relation $\sigma_{xx}=\mu ne$ near the Dirac point \cite{Liu2012}. The charged impurity scattering rate and vacancy scattering rate remain constant with temperature while the phonon scattering rate increases linearly with temperature, as shown in Eqs. (21)-(23). Thus at low temperature and high impurity concentration, $\tau_m$ does not contain significant contributions from phonon scattering, and hence the conductivity and mobility are effectively independent of temperature. The mobility remains almost constant with temperature up to about 50 K, after which it decreases. A similar trend has also been observed in experiments conducted by Chen et al.\ \cite{Chen2008} and Zhu et al. \cite{Zhu_2009} for single-layer graphene. Thus, for samples with high impurity concentrations, impurity scattering dominates the determination of conductivity  at low temperatures while phonon scattering dominates at high temperatures. Conversely, for samples with low impurity concentrations such as highly oriented pyrolitic graphite studied by Sugihara et al.\ \cite{Sugihara1979}, the mobility decreases almost linearly with temperature even at temperatures below 50 K. \par

The room temperature value of conductivity is important for many practical applications; therefore the conductivity at 300 K has been evaluated and shown in Figure A1 for Fermi energy 0.1 eV. The charged impurity and vacancy concentrations have been kept equal and varied along the $x$ axis. The room temperature conductivity remains constant at 30$e^2/h$ for charged impurity concentrations up to $10^{10}$ cm$^{-2}$, after which it decays. This result implies that the room temperature conductivity is primarily influenced by phonon scattering if the concentration of impurities is below a threshold value. Furthermore, the charged impurity and the vacancy concentrations have been varied separately, and the results are tabulated in the inset of Figure A1. An increase in the value of vacancy concentration at a fixed charged impurity concentration leads to a greater reduction in room temperature conductivity compared to an equivalent increase in the value of charged impurity concentration at a fixed vacancy concentration, as suggested from Eqs. (21) and (23). 
\par
Although the Boltzmann transport formalism predicts the electrical conductivity to fall to zero at the Dirac point ($E_F$=0), experimentally, it is found to have a non-zero $\sigma_{min}$. Tan et al.\ \cite{Tan2007} reports that almost all measured $\sigma_{min}$ lie in the range of 2-12$e^2/h$. They justify this observation by the fact that in the low carrier density limit near the Dirac point where the carrier concentration becomes smaller than the charged impurity density, the system breaks up into puddles of electrons and holes where a duality in two dimensions guarantees that, locally, transport occurs either through the hole channel or the electron channel. Most samples are reported to have $\sigma_{min}\approx 4e^2/h$ \cite{Adam2007}. Because we have no way to include such physics in our model, the value of $\sigma_{xx}$ is artificially raised by a value of $4e^2/h$ in the carrier density range $|n|<|n_{imp}|$. 

Figure 2(a) shows the variation of $\sigma_{xx}$ with Fermi energy $E_F$ for different impurity concentrations and then compares it with the experimental results of Liu et al. \cite{Liu2012}. $\sigma_{xx}$ varies symmetrically around the Dirac point both in the model and experimental results, although unlike the experimental results, the model values deviate from parabolic behavior at high $E_F$. \par
Next, we evaluate the influence of magnetic field on the conductivity values.  At non-zero magnetic field, the Lorentz force leads to bending trajectories of thermally diffusing carriers. At zero magnetic field, Lorentz force is absent and both the Hall conductivity $\sigma_{xy}$ and Nernst coefficient $N$ are zero. Figure 2(b) provides comparison of longitudinal conductivity $\sigma_{xx}$ and transverse (Hall) conductivity $\sigma_{xy}$ for zero and non-zero magnetic fields. At zero magnetic field, $\sigma_{xy}$ is zero and $\sigma_{xx}$ varies symmetrically with Fermi energy $E_F$ around the Dirac point. At non-zero magnetic field, $\sigma_{xy}$  behaves as an odd function of $E_F$. Also, $\sigma_{xx}$ loses its symmetry with $E_F$. Such a variation can be inferred from Eq. (28). The second term in Eq. (28) represents the effect of magnetic field. In case of positive $B$ and positive $E_F$, this term is negative. On the other hand, $\sigma_{xy}$ is positive under these conditions. Thus, an increase in $\sigma_{xy}$ corresponds to a decrease in the value of $\sigma_{xx}$. The same trend is observed in experiments conducted by Cho et al. \cite{Cho2008}, shown in Figure 2(b). At a magnetic field of 8 T, oscillations are observed in their experimental results due to the formation of discrete Landau levels. Such oscillations do not appear in our model results because a continuous bandstructure has been assumed. \par

The expressions for $\alpha_{xx}$ and $\alpha_{xy}$ shown in Eqs. (30) and (31) respectively explain certain characteristics about the variation of the Seebeck coefficient $S_{xx}$ and the Nernst coefficient $N$. At zero magnetic field, $\sigma_{xy}$ and $\alpha_{xy}$ are zero. Hence from Eqs. (4) and (5), $N$ is zero and $S_{xx}$ is simply $\alpha_{xx}/\sigma_{xx}$.  Furthermore, $\alpha_{xx}$ and $\alpha_{xy}$ are odd and even functions of $E_F$ respectively such that $S_{xx}$ and $N$ are odd and even functions of $E_F$. The model results for $S_{xx}$ and $N$ obtained from Eqs.(4),(5), and (28)-(31) are compared to available experimental results for one of the samples (mobility 12900  cm$^2$/(V.s)) of Liu et al. \cite{Liu2012} for validation in Figure 3.\par
In the experiment,  $S_{xx}$ was found to be positive below the Dirac point (holes as charge carriers), zero at the Dirac point, and negative above the Dirac point (electrons as charge carriers). Similar trends are observed in the model results plotted in the same figure. As $E_F$ increases from zero, the magnitude of $S_{xx}$ increases, reaches a maximum, and then decreases. Impurity and vacancy concentrations of 10$^{11}$ cm$^{-2}$ provide the best match with the experimental results.  \par
For both experiment and model, the Nernst coefficient $N$ reaches its peak value near zero $E_F$ and decays slowly with increasing Fermi energy $E_F$ to a negative value before rising back to a positive value at very high $E_F$. Close to the Dirac point, the electrons and holes produce an additive effect on the Nernst coefficient as they are deflected in opposite directions upon applying a magnetic field and thereby add to the transverse voltage. At high $E_F$, the asymptotic decay is predicted by Fermi liquid theory according to which $N$ vanishes due to Sondheimer's cancellation \cite{sharma2015,Wang2001}. Comparing model with experiment, the decay in $N$ upon moving away from zero $E_F$ is observed to be steeper in the experiment than in the model. Impurity and vacancy densities of 2.5$\times 10^{11}$ cm$^{-2}$ provide the best match with experimental results.\par
Thus to summarize, the variation of the Seebeck and Nernst coefficients with Fermi energy qualitatively matches those obtained from past experiments, although a single impurity density is not able to match the peak values of both $S_{xx}$ and $N$.  

\par

\subsection{Nernst coefficient}
In this section, the influence of impurity concentration and temperature on the Nernst coefficient is studied. The magnetic field is considered to be 1 T because we have neglected higher order terms in $B$ in the transport coefficient calculations. At high magnetic fields, the electronic band structure breaks down into discrete Landau levels and such effects are not considered in our analysis. Figure 4 shows variation of the Nernst coefficient $N$ with Fermi energy for graphene samples with different impurity concentrations at 300 K. The peak value of $N$ decreases with increasing impurity concentration. The peak values are 50, 20, 11 and 7 $\mu$VK$^{-1}$T$^{-1}$for impurity concentrations of 2.5 $\times$ 10$^{11}$ cm$^{-2}$, 5 $\times$ 10$^{11}$ cm$^{-2}$, 7.5 $\times$ 10$^{11}$ cm$^{-2}$ and 10$^{12}$ cm$^{-2}$ respectively. Moreover, as the impurity concentration decreases, the full width at half maximum (FWHM) decreases. The results are consistent with the experimental resuts of Liu et al.\ \cite{Liu2012} who found the Nernst signal $S_{xy}$ to increase almost linearly with mobility. Mobility of devices decreases with increased impurity concentration as shown in Figure 1(b) and (2).

The only study on the temperature dependence of $N$ is that of Wei et al.\ \cite{Wei2009}. Their experiments on mechanically exfoliated graphene showed a reduction in oscillations of $N$ and a rise in peak values with increasing temperature. A rigorous analysis of the influence of temperature needs to be conducted. Figure 5 shows the variation in $N$ from 10 to 400 K for samples with impurity concentrations of $2.5\times 10^{11}$ cm$^{-2}$ and 5$\times$10$^{11}$ cm$^{-2}$. The shapes of the curves change slightly with increasing temperature. At low temperatures, the curve is quite flat, and $N$ remains positive for almost all values of $E_F$. But as temperature increases, at low $E_F$ $N$ attains higher positive peaks. In Figure 5(a), the peak value rises from 5 to 54 $\mu$V K$^{-1}$T$^{-1}$ upon varying temperature from 10 to 200 K and then slightly decreases to 40 $\mu$V K $^{-1} $T$^{-1}$) at 400 K. On the other hand, in Figure 5(b), $N$ increases from 1 to 20 $\mu$VK$^{-1}$T$^{-1}$from 10 to 300 K and then declines to 19 $\mu$VK$^{-1}$T$^{-1}$at 400 K. The temperature dependence of $N$ is shown for different values of impurity concentrations at a fixed $E_F$ of 0.05 eV in Figure 6(a). At the lowest impurity concentration, $N$ increases rapidly from 4 to 37 $\mu$V K$^{-1}$T$^{-1}$ after which it decreases slightly to 28 $\mu$V K$^{-1}$T$^{-1}$ at 400 K. For higher impurity concentrations, $N$ increases at a slower pace with temperature until 300 K, after which it saturates. The trends in Figure 6(b) showing $N$'s variation with temperature for different values of Fermi energy at a fixed impurity concentration are similar.

\par
Figure 7 displays variations in $N$ with Fermi energy for different values of band gap $\bigtriangleup$. Introducing a band gap leads to a decrease in the peak value of $N$. The peak value decreases from 50 $\mu$V K$^{-1}$T$^{-1}$ at $\bigtriangleup$=0 to 13 $\mu$VK$^{-1}$T$^{-1}$at $\bigtriangleup$=100 meV. Also, the curves are found to peak at Fermi energies near the bottom of the conduction band. This can be explained by the fact that for Fermi energy in the range of -$\bigtriangleup$ to $\bigtriangleup$, the density of states of both electrons and holes are zero which implies relatively few carriers undergoing deflection due to magnetic field. 

\par
\subsection{Seebeck coefficient}
Figure 8 shows the Seebeck coefficient $S_{xx}$ as a function of Fermi Energy $E_F$ for various charged impurity and vacancy concentrations. The peak values of $S_{xx}$ are lower for higher concentrations. They decrease from 60 $\mu$V/K at 10$^{11}$ cm$^{-2}$ to 46 $\mu$V/K at 10$^{12}$ cm$^{-2}$. Furthermore at high impurity concentrations, the curves are found to exhibit a smoother transition (over a larger $E_F$) from positive to negative near the Dirac point. These trends are consistent with experimental results \cite{Liu2012} as well as with tight binding model results \cite{2010_Hao}. Liu et al.\ \cite{Liu2012} observed that $S_{xx}$ increases from ~50 to 75 $\mu$V/K with increased mobility from 4560 to 12900 cm$^2$/(V.s). Hao et al.\ \cite{2010_Hao} found that as the impurity density is raised from 0 to 5$\times$ 10$^{12}$ cm$^{-2}$, the peak Seebeck coefficient is found to occur at higher values of Fermi energy. 

\par
Figure 9 shows the Seebeck coefficient $S_{xx}$ as a function of Fermi Energy $E_F$ for different operating temperatures. The impurity concentration is fixed at 2.5 $\times 10^{11}$ cm$^{-2}$. As temperature increases, $S_{xx}$ increases monotonically until 400 K at low $E_F$. However it reaches a peak and decreases slightly near the Dirac point. The inset of Figure 9 shows the variation of the peak value $|S_{xx}|_{max}$ with temperature. The peak increases from 10 to 300 K after which it decreases slightly. 

\par
The foregoing results suggest that the Seebeck coefficient is limited to around 60 $\mu$V/K at zero magnetic field, as also been confirmed experimentally. We now consider the possibility of accentuating the Seebeck coefficient $S_{xx}$ by applying a perpendicular magnetic field or introducing a band gap. Figure 10 shows the variation of $S_{xx}$ with Fermi energy $E_F$ at different values of magnetic field, from -1 to +1 T. The curves become asymmetric upon applying a magnetic field although they all cross zero at the Dirac point. At negative magnetic field, the peak is ~71 $\mu$V/K in the negative Fermi energy region and ~-54 $\mu$V/K in the positive region. The trend reverses upon changing the direction of magnetic field.Thus applying a magnetic field of 1 T can accentuate the peak Seebeck coefficient from 60 to 71 $\mu$V/K. 
\par
Previous experimental \cite{2011_Xiao} and theoretical studies \cite{Patel2012} have shown that the introduction of a band gap accentuates the Seebeck coefficient $S_{xx}$. Figures 11 and 12 show the effect of band gap from 0 to 100 meV at zero magnetic field. In Figure 11, $S_{xx}$ is observed to rise with increase band gap at all $E_F$ values. The inset shows similar trends obtained using Green's function formalism for graphene nanoribbons \cite{2011_Mazzamuto}. The shape of the curves is similar to that of our study. Furthermore, their peak value predictions of 110 and 280 $\mu$V/K for band gaps of 56 meV and 105 meV respectively, are close to our results of 161 and 273 $\mu$V/K for 50 meV and 100 meV band gaps respectively. Figure 12 indicates that the peak value of $S_{xx}$ increases linearly with band gap. This result is expected because upon introduction of a band gap, the electrons and holes no longer have nullify each other near the band edge, thus producing a higher value of $S_{xx}$. The linear relationship between the maximum Seebeck coefficient and band gap conforms with the theoretical prediction of Goldsmid et al. \cite{1999_Goldsmid} for a general semiconductor. The inset of Figure 12 shows the Fermi energy $E_F$ corresponding to the peak values. It starts at 45 meV for zero band gap, falls to 41 meV at a band gap of 25 meV, and thereafter rises linearly to 64 meV at a band gap of 100 meV. The underlying reason may be that the band edge shifts to a higher Fermi energy as band gap increases.

\section{Conclusions}
In summary, an analytical solution to the BTE has been used to derive the electric and thermoelectric transport coefficients of single-layer periodic graphene both in the presence and absence of magnetic field. By incorporating phonon, charged impurity and vacancy scattering mechanisms into the analysis, we are able to replicate experimentally measured results qualitatively, and also quantitatively to a certain degree. Impurity concentration is found to be a critical factor in determining the thermoelectric and thermomagnetic transport coefficients. The results include specific predictions for the Seebeck and Nernst coefficients, some of which need to be verified experimentlally. At high impurity concentrations, the Nernst coefficient has a lower peak around zero Fermi energy and shows less variation at high Fermi energy. Similarly, the Seebeck coefficient has a higher peak value for low impurity concentrations. Upon increasing temperature, the peak Nernst coefficent near the Dirac point increases monotonically to 300 K, following which it decreases slightly. Creating a band gap in the bandstructure enhances the Seebeck coefficient but degrades the Nernst coefficient. Detailed knowledge of the variation of the thermoelectric and thermomagnetic properties of graphene shown in this paper may be helpful for improved magnetothermoelectric coolers and sensors. 
\\
\FloatBarrier
\clearpage

\section*{Appendix}

Sample derivation of longitudinal electrical conductivity: Eq. (11) for $\sigma_{xx}$ in the absence of magnetic field becomes
\begin{equation}\label{eq31}
\begin{split}
\sigma_{xx} &=4e^2\int_{-\infty}^{\infty}\frac{|2\pi kdk|}{(2\pi)^2}\frac{v_F^2}{2}\tau_m(-\frac{\partial f_{eq}}{\partial \epsilon})  \\
&=4e^2\int_{-\infty}^{0}\frac{|2\pi kdk|}{(2\pi)^2}\frac{v_F^2}{2}\tau_m(-\frac{\partial f_{eq}}{\partial \epsilon})  \\
& +4e^2\int_{0}^{\infty}\frac{|2\pi kdk|}{(2\pi)^2}\frac{v_F^2}{2}\tau_m(-\frac{\partial f_{eq}}{\partial \epsilon}) \\
&= \sigma_{xx,+}+\sigma_{xx,-}
\end{split}
\end{equation}
\\
\\ \\ \\ \\
For $\sigma_{xx,+}$
\begin{equation} \label{eq33}
\begin{split}
-\frac{\partial f_{eq}}{\partial \epsilon}&=\frac{\frac{1}{k_bT}exp(\frac{\epsilon-\mu}{k_bT})}{(1+exp(\frac{\epsilon-\mu}{k_bT}))^2}\\ & =\frac{1}{k_bT(exp(\frac{\epsilon-\mu}{k_bT})+2+exp(-\frac{\epsilon-\mu}{k_bT}))} \\
&=\frac{1}{k_bT(2+2cosh(\frac{\epsilon-\mu}{k_bT}))} \\& =\frac{1}{k_bT(4cosh^2(\frac{\epsilon-\mu}{2k_bT}))}
\end{split}
\end{equation}

For $\sigma_{xx,-}$, performing the variable transform  $\epsilon'=-\epsilon$
\begin{equation} \label{eq34}
-\frac{\partial f_{eq}}{\partial \epsilon}=\frac{1}{k_bT(4cosh^2(\frac{\epsilon+\mu}{2k_bT}))}
\end{equation}
\begin{equation}  \label{eq35}
-\frac{\partial f_{eq}}{\partial \epsilon}=\frac{1}{k_bT(4cosh^2(\frac{\epsilon+\mu}{2k_bT}))}
\end{equation}
Considering the expression for $\tau_m$ in Eq. (25), and with the variable transform $x=E/E_F$, we obtain the final expression for $\sigma_{xx}$ (Eq. (28). For a band gap $\bigtriangleup$, the expression for $\sigma_{xx}$ is modified using Eqs. (26), (27) to obtain
\begin{widetext}
\begin{equation}
\begin{split} \label{eq37}
\sigma_{xx}=&\frac{2e^2}{h}\frac{E_F^2|k_F|}{4\hbar k_bT}\int_{\frac{\bigtriangleup}{E_F}}^\infty (1-\frac{\bigtriangleup^2}{x^2E_F^2})\frac{x^2(cosh^{-2}(\frac{E_F}{2k_bT}(x-1))+cosh^{-2}(\frac{E_F}{2k_bT}(x+1)))}{\alpha_1+\alpha_2k_F^2x^2+\frac{\alpha_3}{|ln(k_FxR_0)|^2}}dx \\
&-\frac{2e^2}{h}\frac{eBv_F^3k_F^3}{8k_bT}\int_{\frac{\bigtriangleup}{E_F}}^\infty (1-\frac{\bigtriangleup^2}{x^2E_F^2})\frac{x^2(cosh^{-2}(\frac{E_F}{2k_bT}(x-1))-cosh^{-2}(\frac{E_F}{2k_bT}(x+1)))}{(\alpha_1+\alpha_2k_F^2x^2+\frac{\alpha_3}{|ln(k_FxR_0)|^2})^2}.
\end{split}
\end{equation}
\end{widetext}

\newpage

\bibliographystyle{plainnat}
\bibliography{paper}
\clearpage
\newpage
\newpage
\title{Figures}
\begin{figure} 
\includegraphics[width=8.5cm, height=6cm]{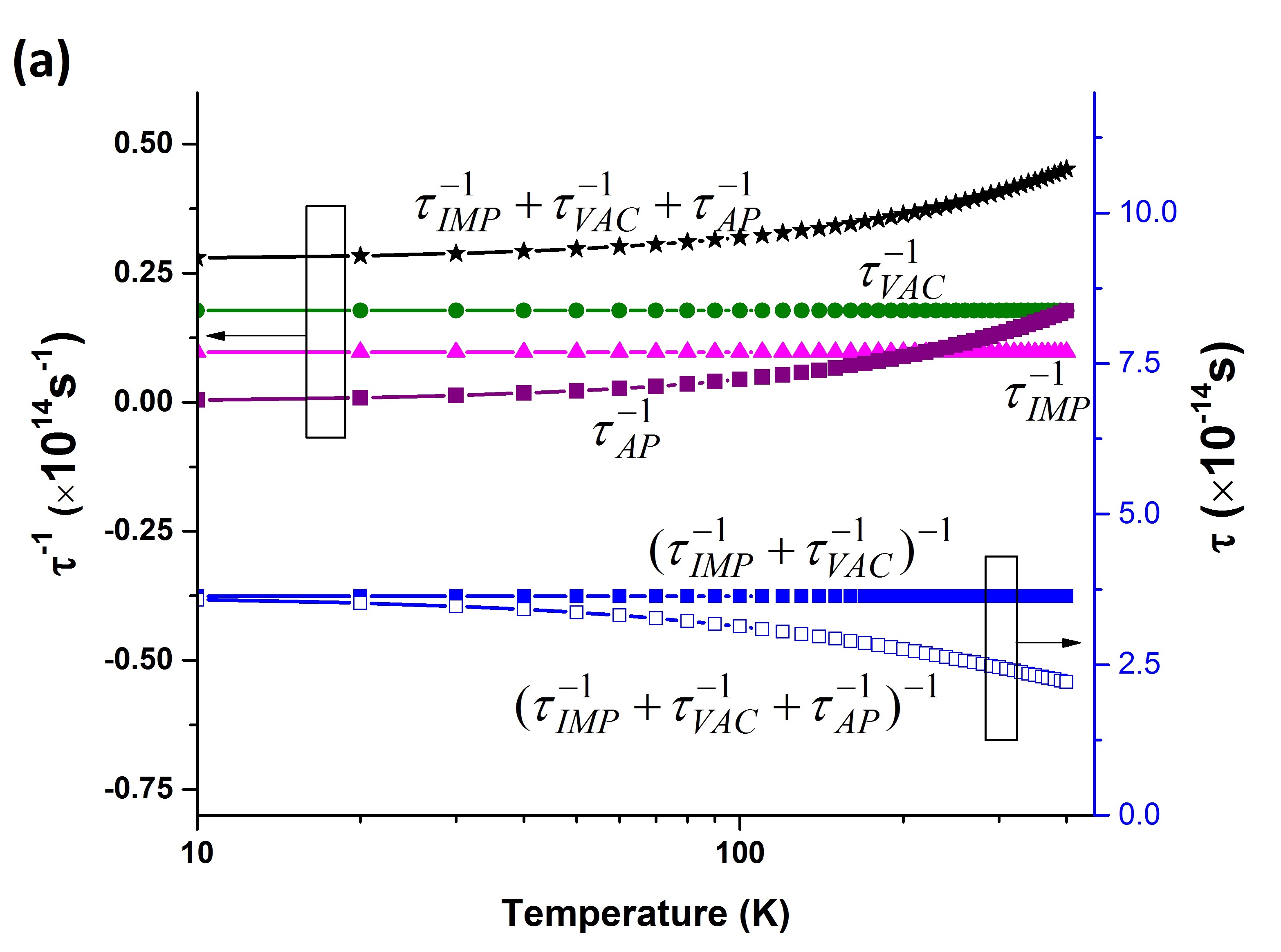}
\includegraphics[width=8.5cm, height=7cm]{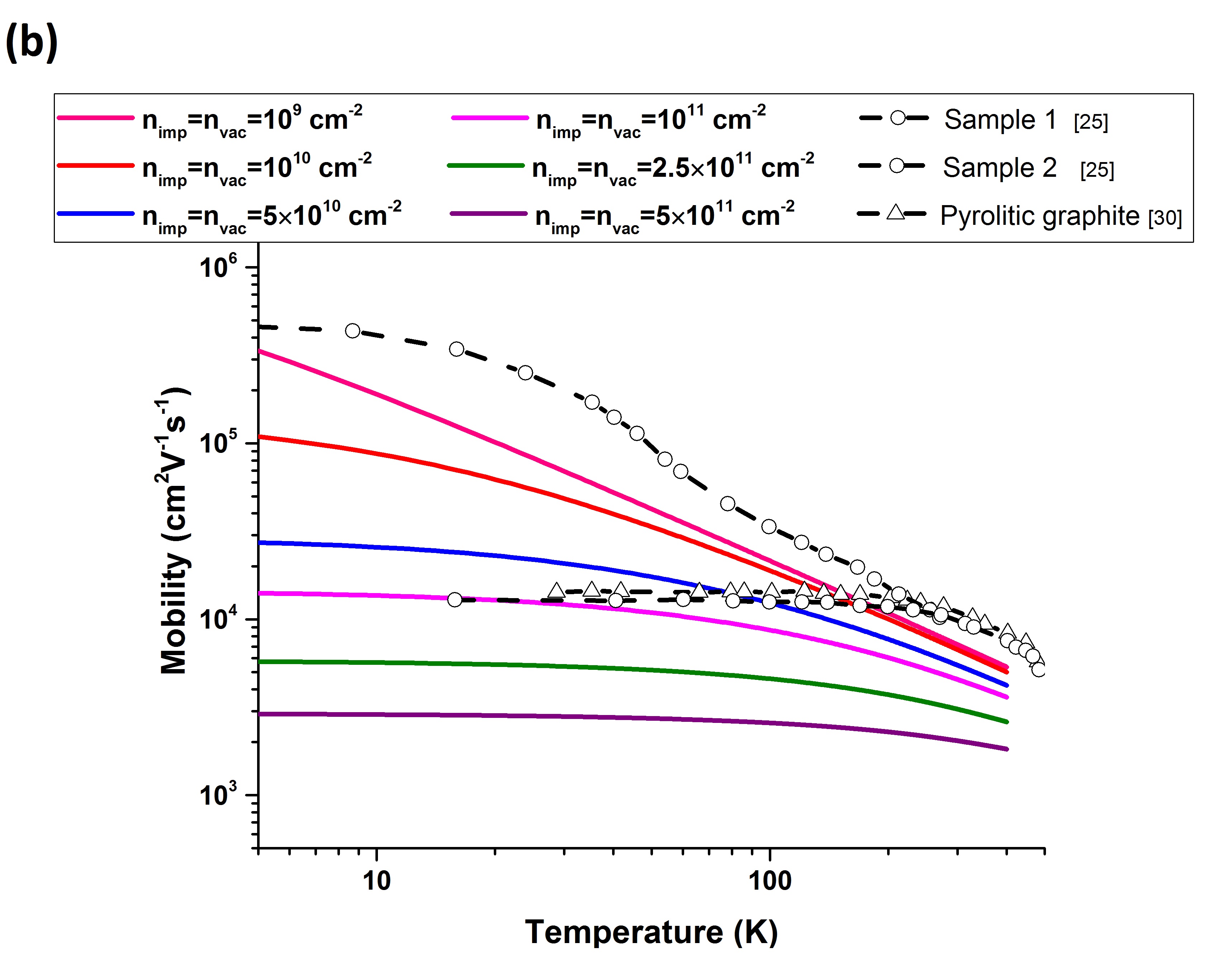}
\label{Fig1}
\caption{(a) Variation of scattering time $\tau$ and scattering rate $\tau^{-1}$ for different scattering mechanisms, with temperature $T$, at charged impurity concentration $n_{imp}$ and vacancy concentration $n_{vac}$ both equal to 5$\times 10^{11}$cm$^{-2}$, and Fermi energy $E_F$=0.1 eV.  (b) Mobility as a function of temperature $T$ for different impurity and vacancy concentrations and comparison with experiments by Chen et al. \cite{Chen2008} and Sugihara et al. \cite{Sugihara1979} at Fermi energy $E_F$=0.1 eV.   } 
\end{figure}

\begin{figure} 
\includegraphics[width=9cm, height=7cm]{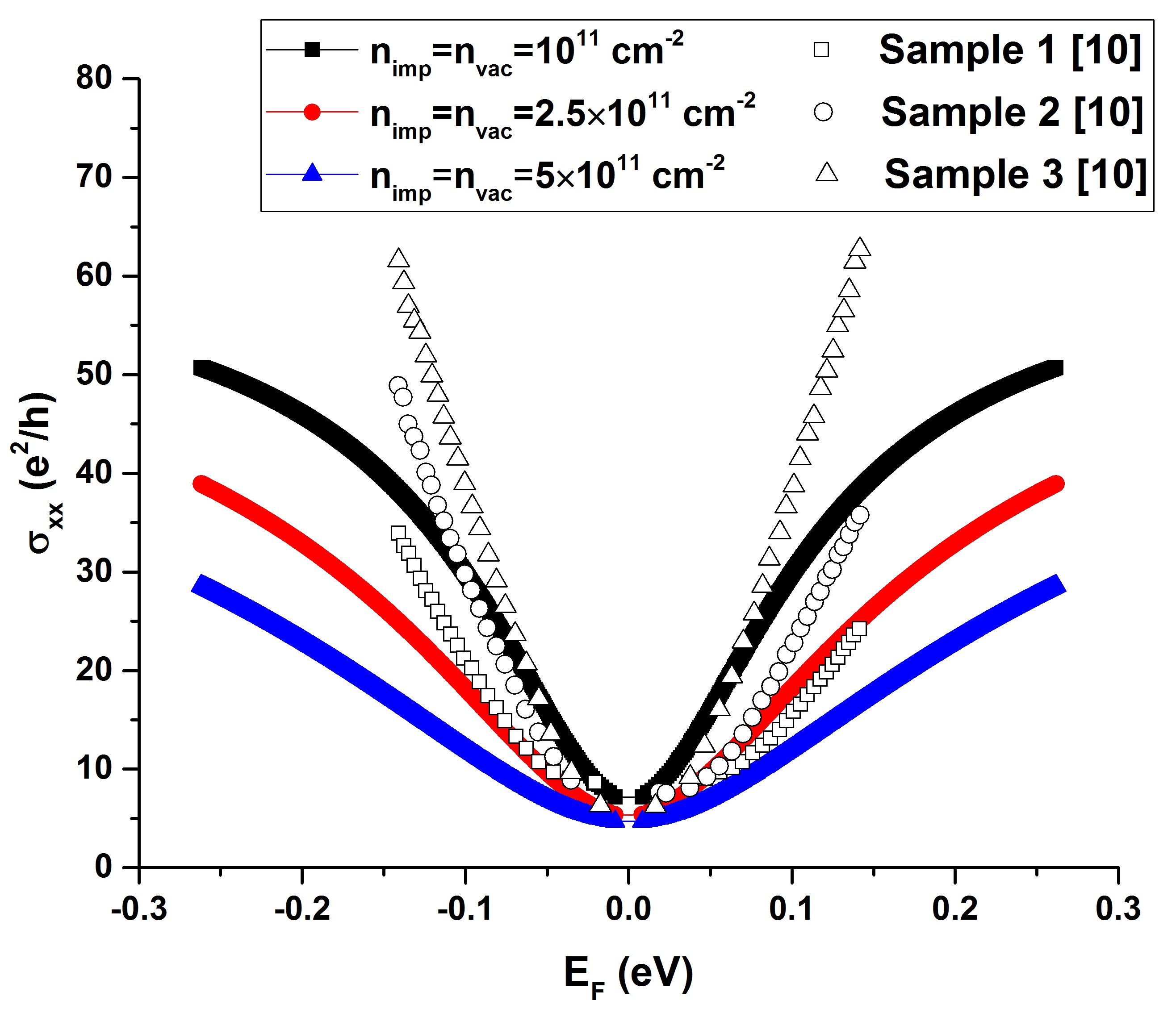}
\includegraphics[width=9cm, height=7cm]{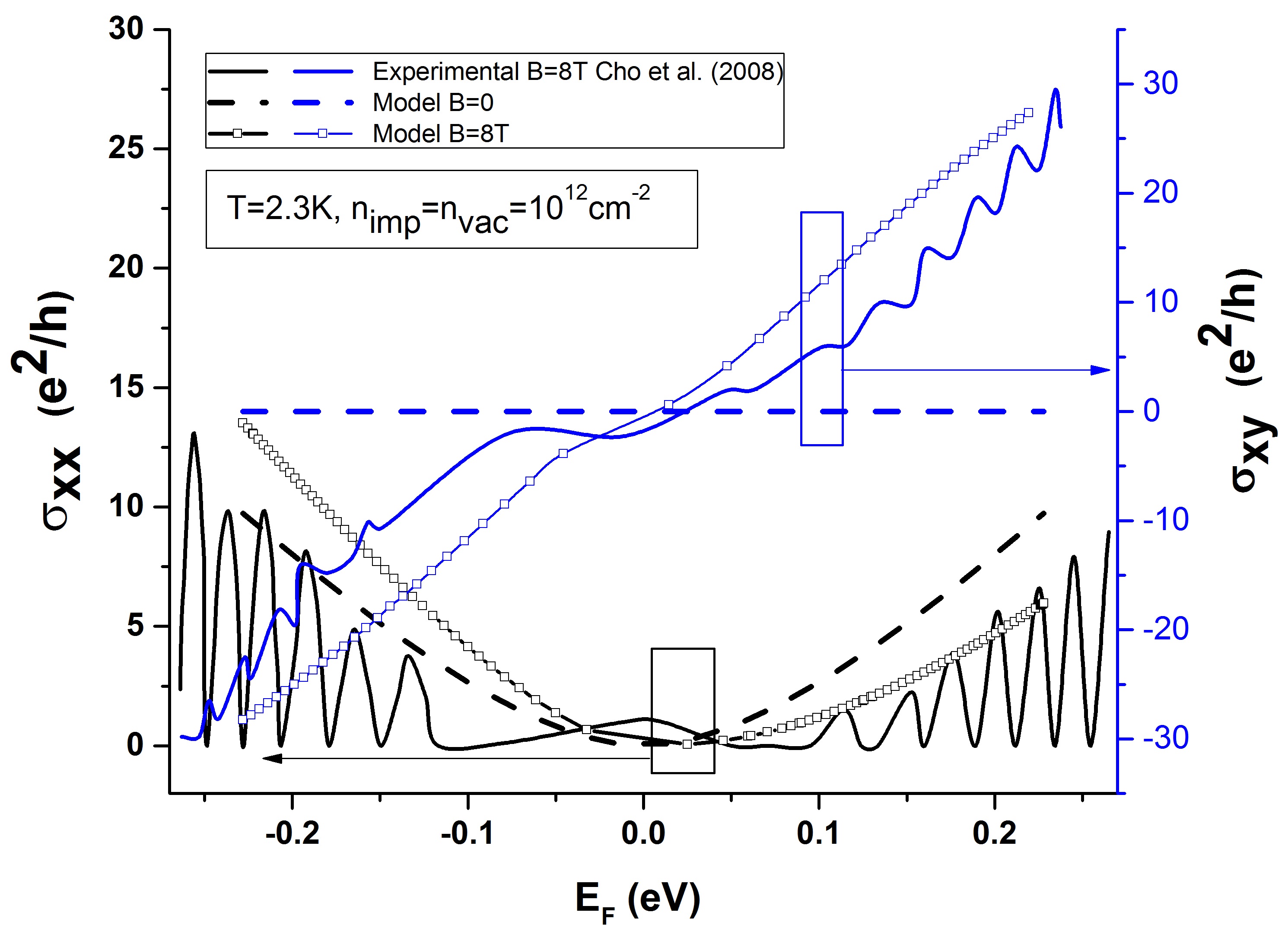}
\label{Fig2}
\caption{(a) Electrical conductivity $\sigma_{xx}$ normalized by $e^2/h$ at temperature $T$=150 K, magnetic field $B$=0 for various charged impurity and vacancy concentrations, and its comparison with experimental results of Liu et al.\ \cite{Liu2012}. (b) Longitudinal conductivity $\sigma_{xx}$ and transverse conductivity $\sigma_{xy}$ obtained from the model compared to experimental results from Cho et al. \cite{Cho2008} $B$=8 T and $T$=2.3 K for both experiment and model. Impurity and vacancy concentrations are chosen be 1$\times$10$^{12}$ cm$^{-2}$ in the model.  } 
\end{figure}

\begin{figure} 
\includegraphics[width=9cm, height=7cm]{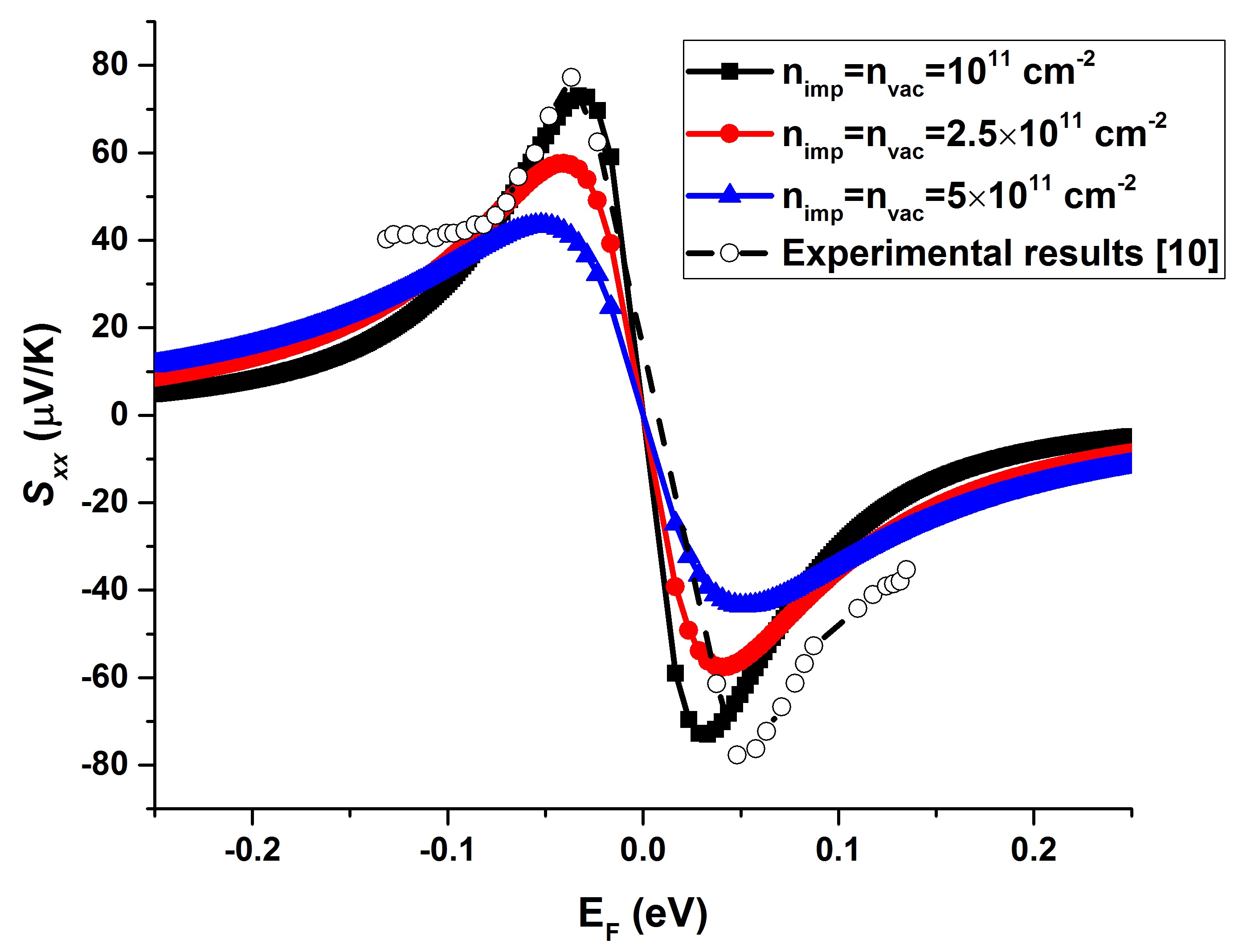}
\includegraphics[width=9cm, height=7cm]{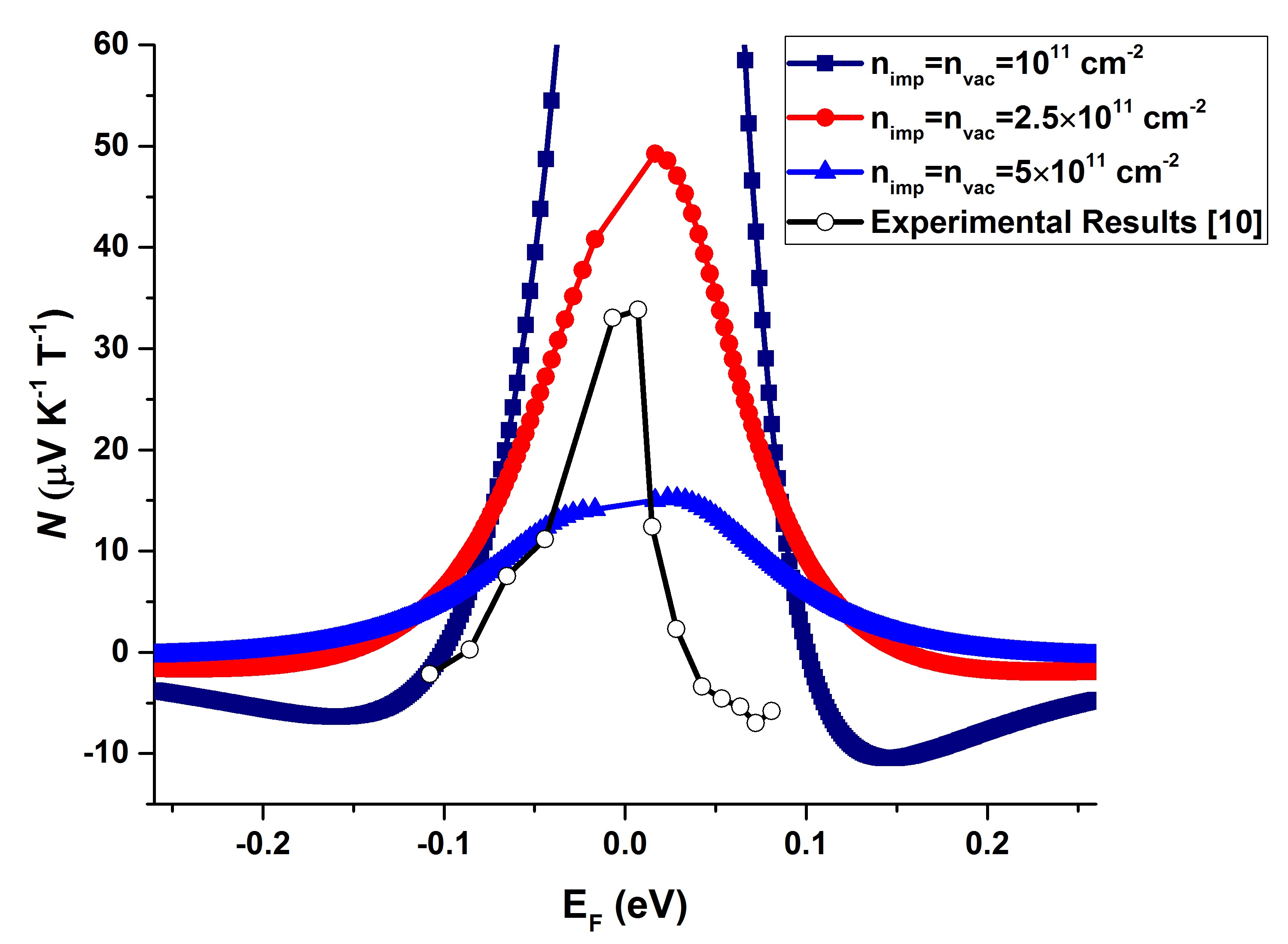}
\label{Fig3}
\caption{Comparison of (a)model Seebeck coefficient S$_{xx}$ and (b)model Nernst coefficient $N$ with experimental results of Liu et al. [10]. Temperature T=150 K and magnetic field B=0 for (a). Temperature T=150 K and magnetic field B=1 T for (b).}
\end{figure}

\begin{figure} 
\includegraphics[width=9cm, height=7cm]{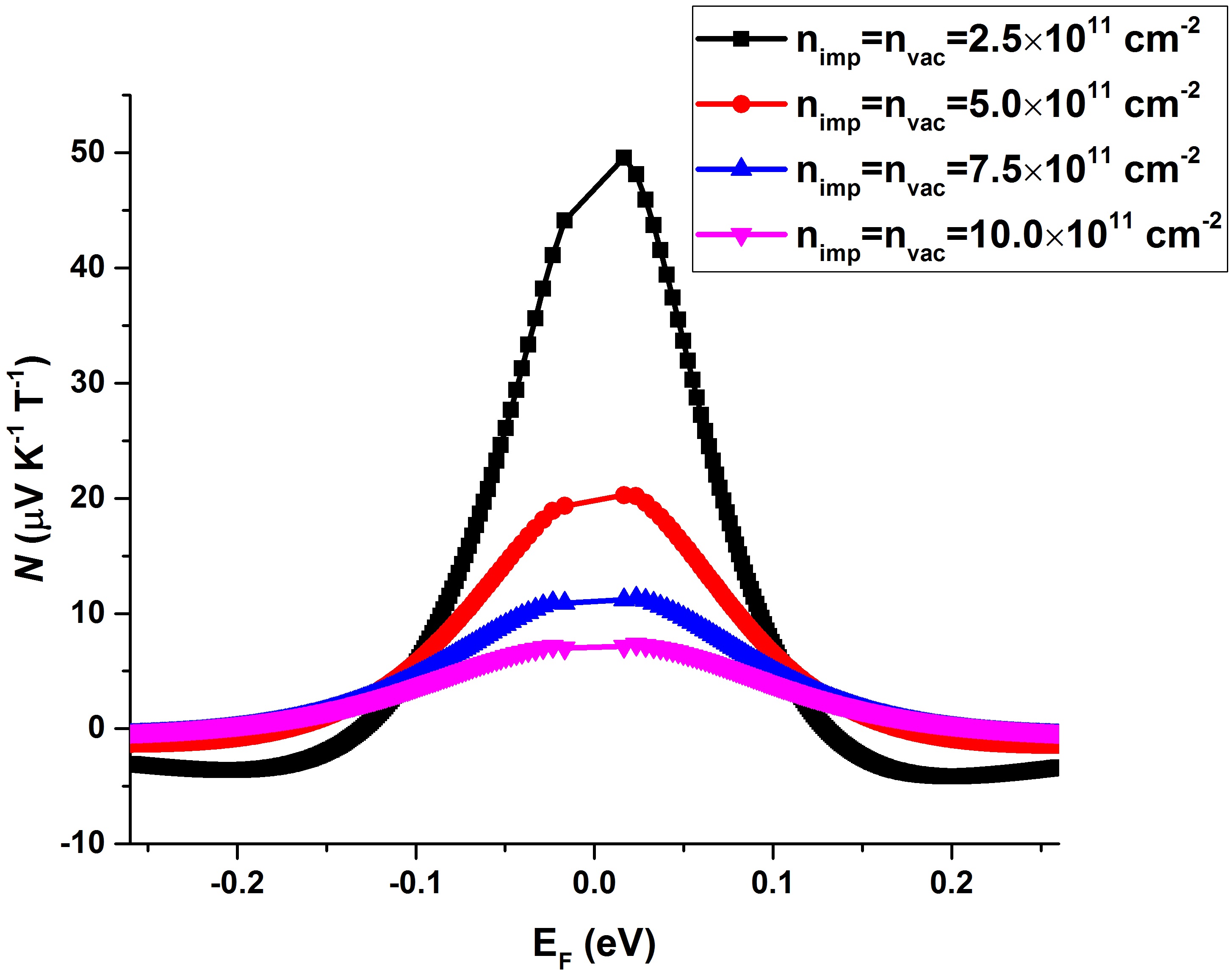}
\label{Fig4}
\caption{ Variation of Nernst coefficient $N$ with charged impurity concentration $n_{imp}$ and vacancy concentration $n_{vac}$ at temperature $T$=300 K and magnetic field $B$=1 T. } 
\end{figure}

\begin{figure} 
\includegraphics[width=9cm, height=7cm]{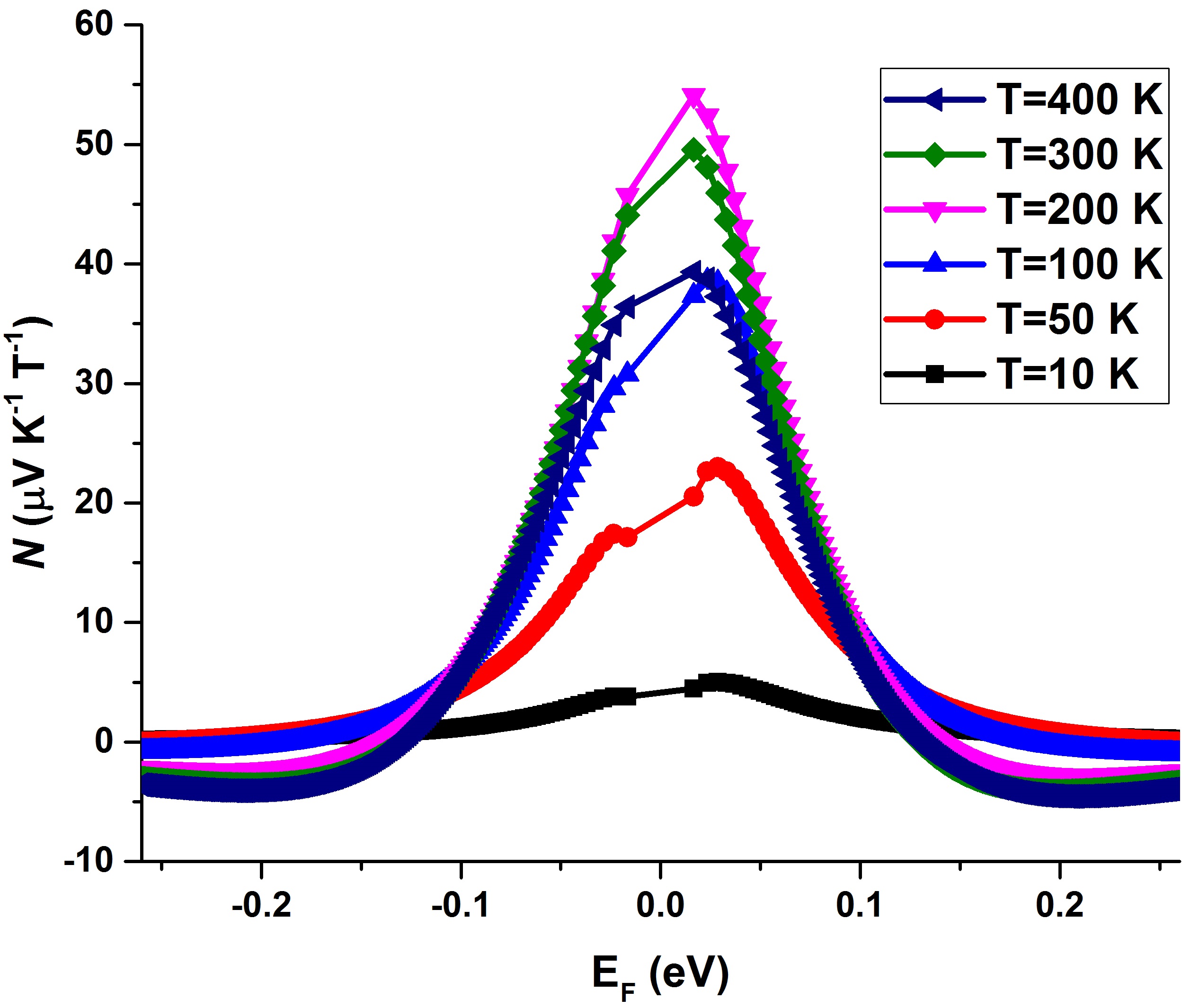}
\includegraphics[width=9cm, height=7cm]{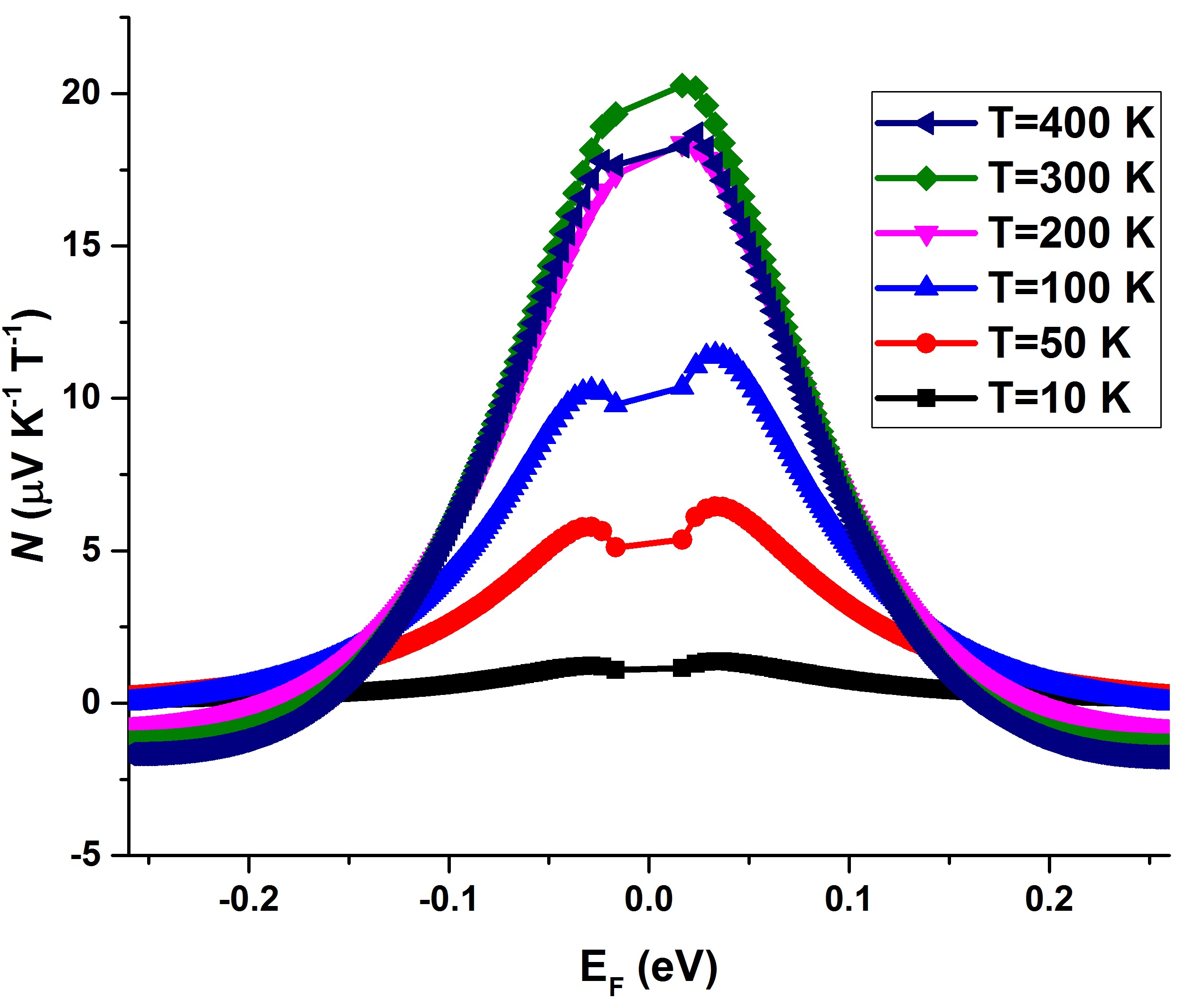}
\label{Fig5}
\caption{Variation of Nernst coefficient $N$ with Fermi energy E$_{F}$ for temperature T varying from 10 K to 400 K. Magnetic field B=1 T, impurity concentration n$_{imp}$ and vacancy
concentration n$_{vac}$ are both equal to (a) 2.5$\times$10$^{11}$ cm$^{-2}$.(b)5$\times$10$^{11}$ cm$^{-2}$.}
\end{figure}

\begin{figure} 
\includegraphics[width=9cm, height=8cm]{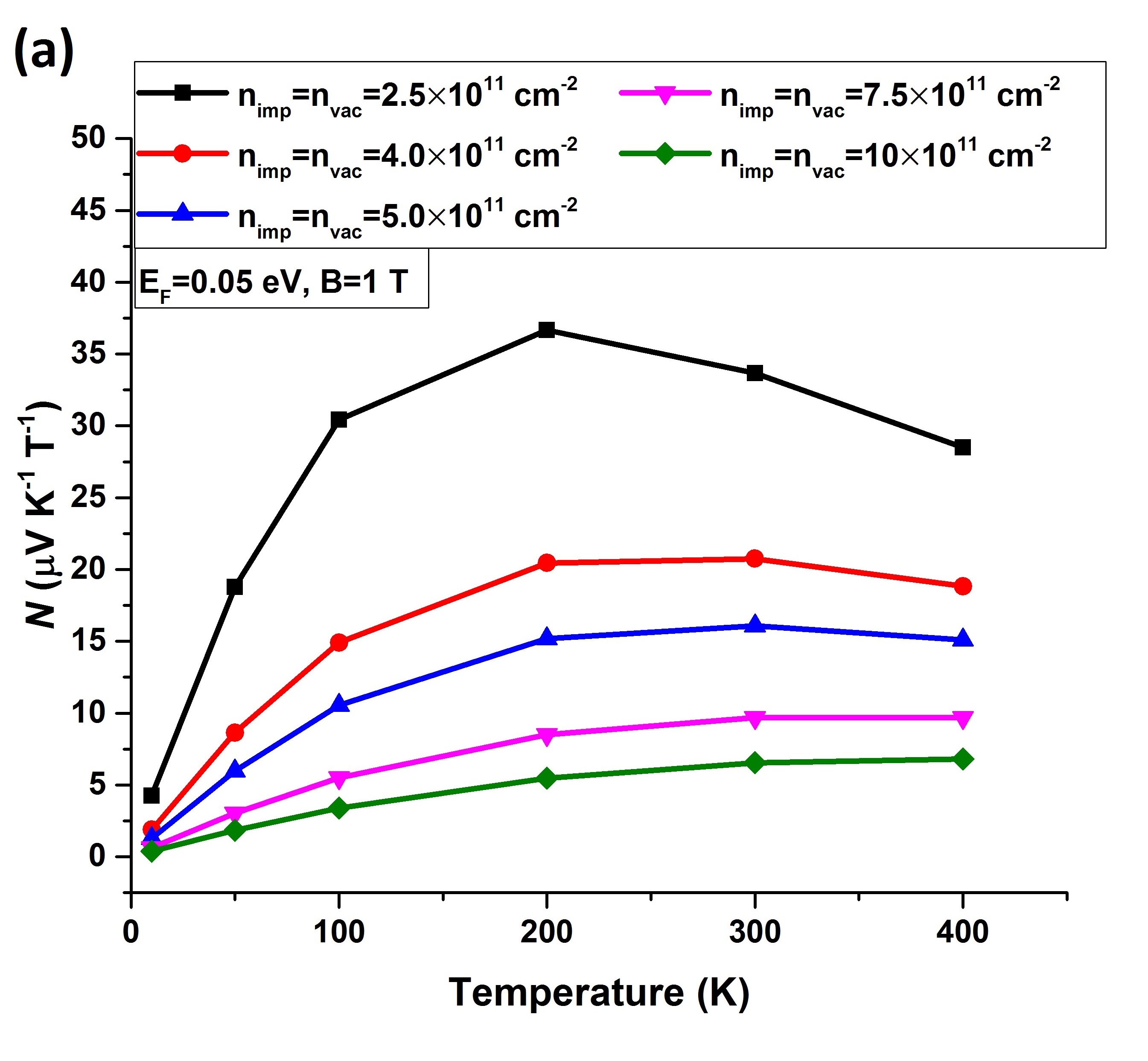}
\includegraphics[width=9cm, height=8cm]{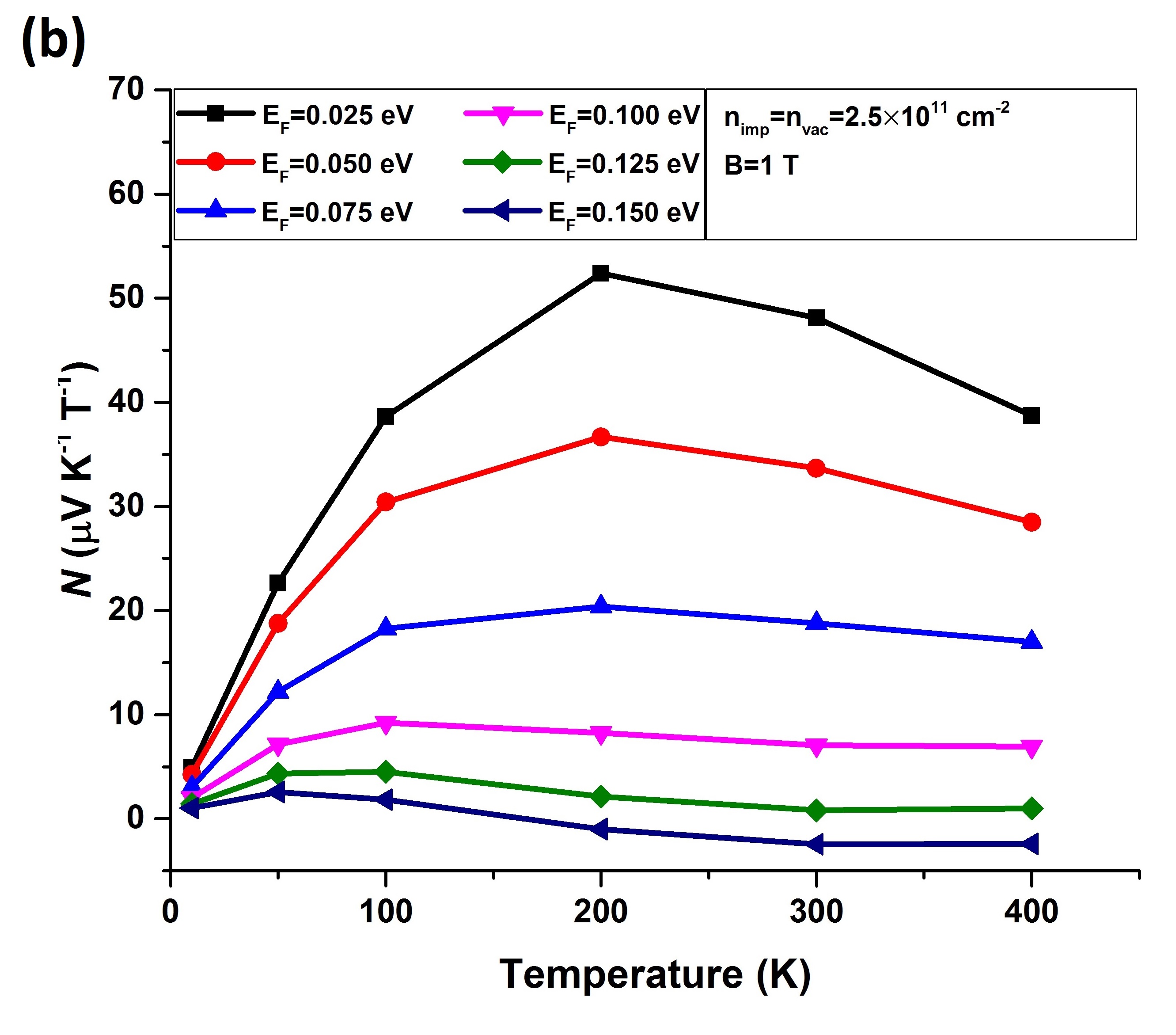}
\label{Fig6}
\caption{(a) Variation of Nernst coefficient $N$ with temperature $T$ for different charged impurity concentration $n_{imp}$ and vacancy concentration $n_{vac}$ at fixed Fermi energy $E_F$= 0.05 eV and magnetic field $B$=1 T. (b) Variation of Nernst coefficient $N$ with temperature $T$ for different values of Fermi energy $E_F$ at fixed charged impurity concentration $n_{imp}$ and vacancy concentration $n_{vac}$ of 2.5$\times 10^{11}$ cm$^{-2}$, and magnetic field $B$=1 T.} 
\end{figure}

\begin{figure} 
\includegraphics[width=8.5cm, height=7cm]{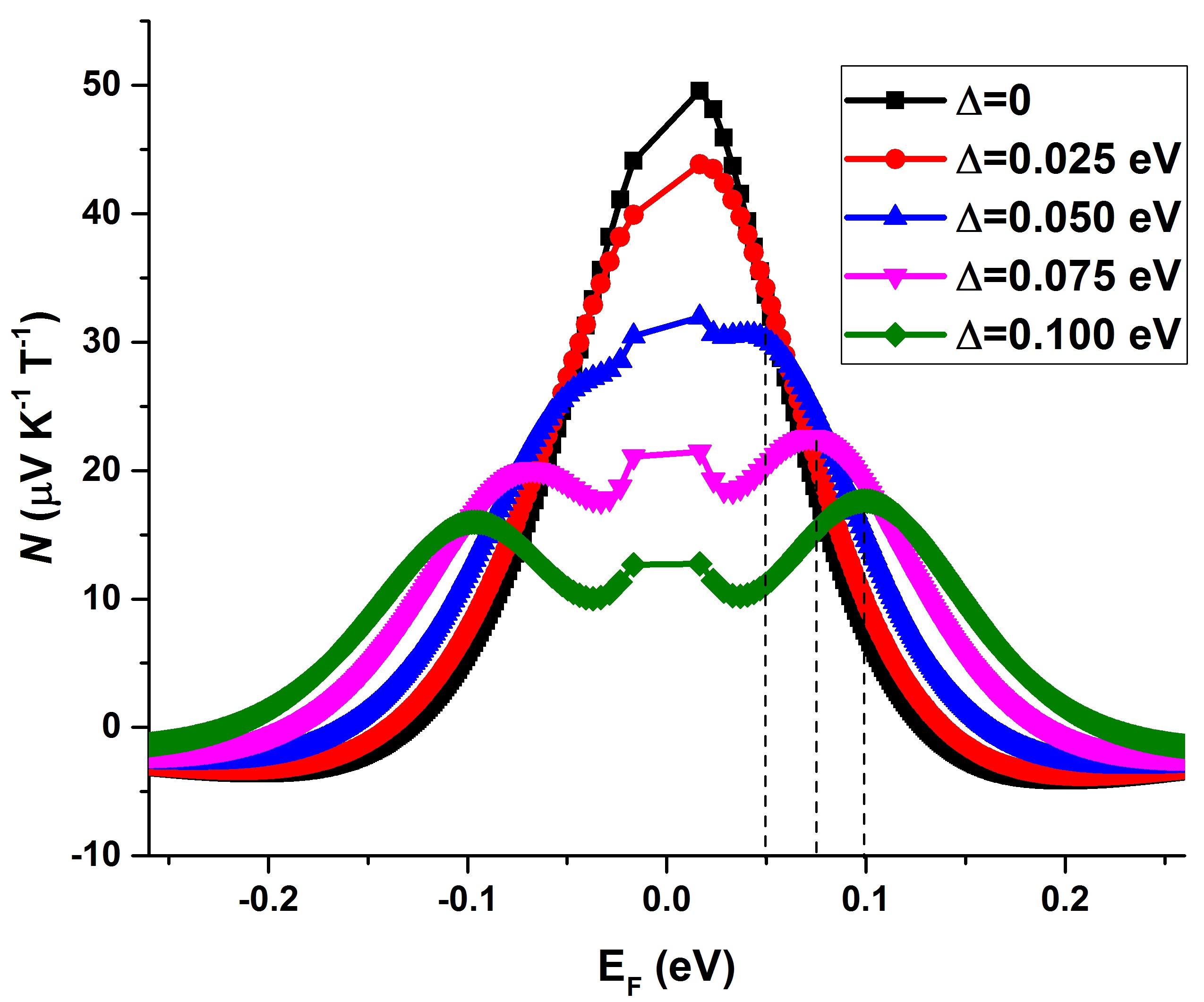}
\label{Fig7}
\caption{Variation of Nernst coefficient $N$ with Fermi energy $E_F$ at magnetic field B=1 T, temperature $T$=300 K, charged impurity concentration $n_{imp}$ and $n_{vac}$ both equal to 2.5$\times$10$^{11}$ cm$^{-2}$.} 
\end{figure}

\begin{figure} 
\includegraphics[width=8.5cm, height=7.5cm]{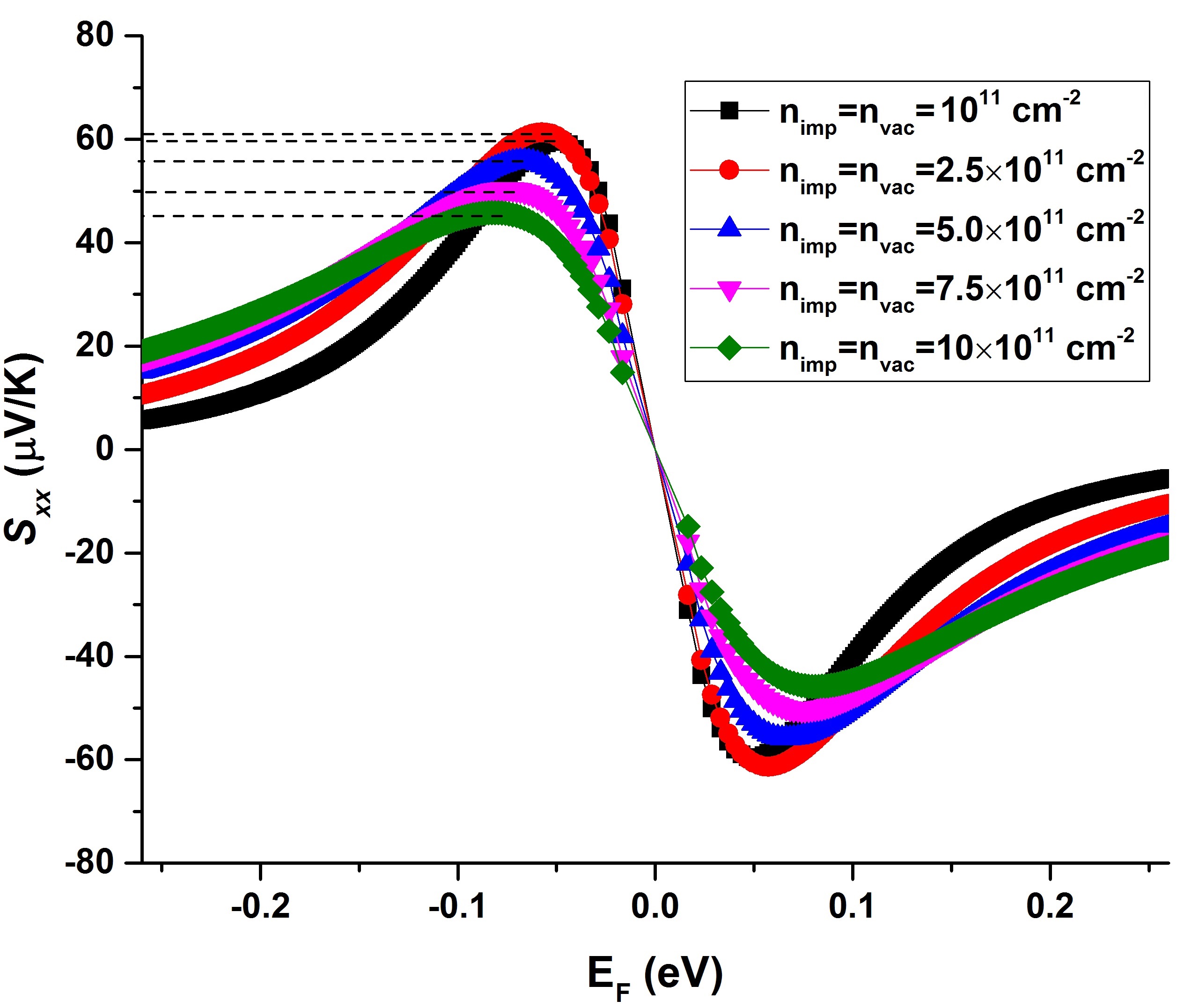}
\label{Fig8}
\caption{Variation of Seebeck coefficient $S_{xx}$ with Fermi energy $E_F$ for different impurity concentrations at temperature $T$= 300 K and magnetic field $B$=0.} 
\end{figure}

\begin{figure} 
\includegraphics[width=9cm, height=8cm]{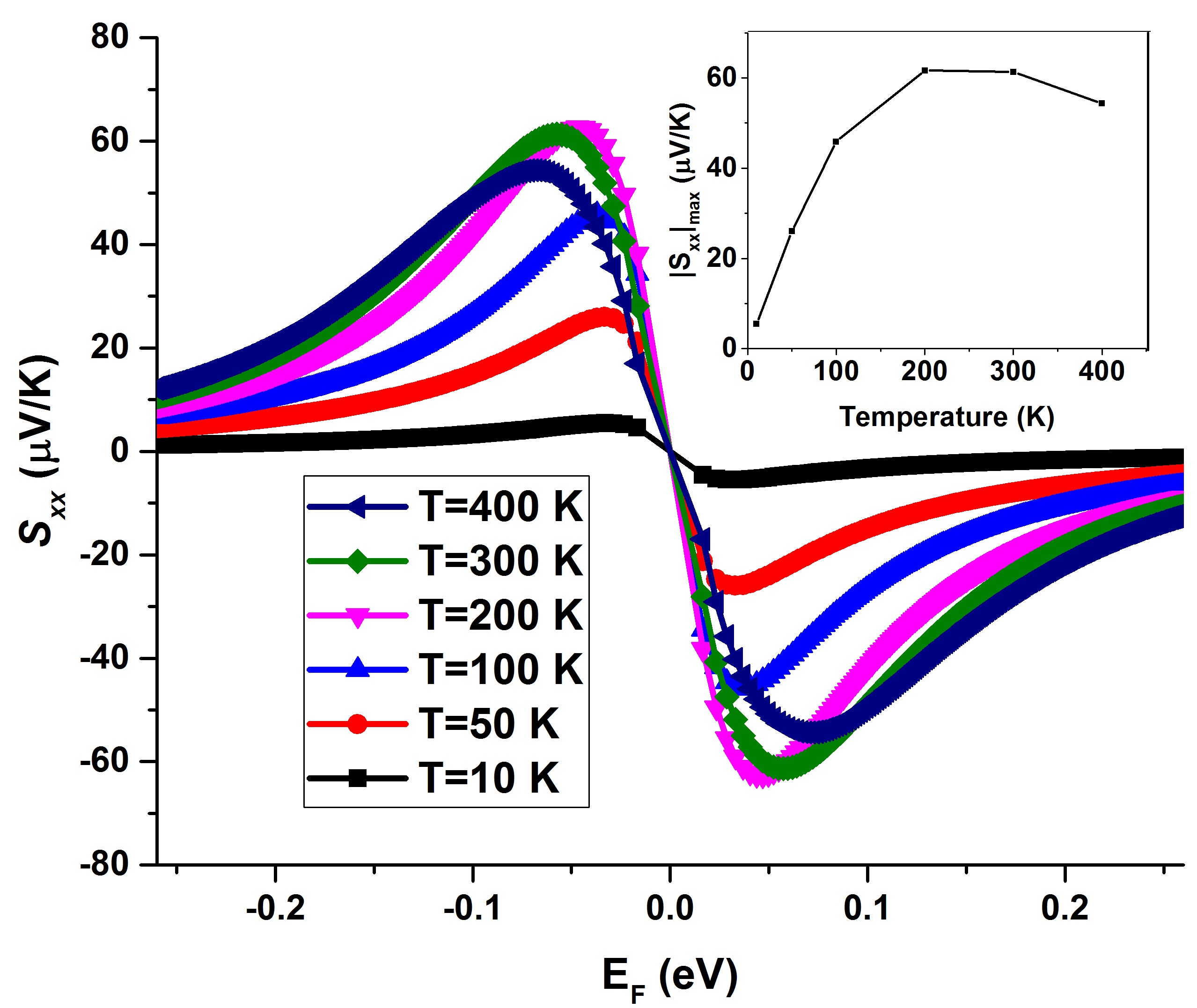}
\label{Fig9}
\caption{Variation of Seebeck coefficient $S_{xx}$ with Fermi energy $E_F$ for different temperatures at $n_{imp}=n_{vac}$=2.5$\times$ 10$^{11}$ cm$^{-2}$ and magnetic field $B$=0. } 
\end{figure}

\begin{figure} 
\includegraphics[width=9cm, height=8cm]{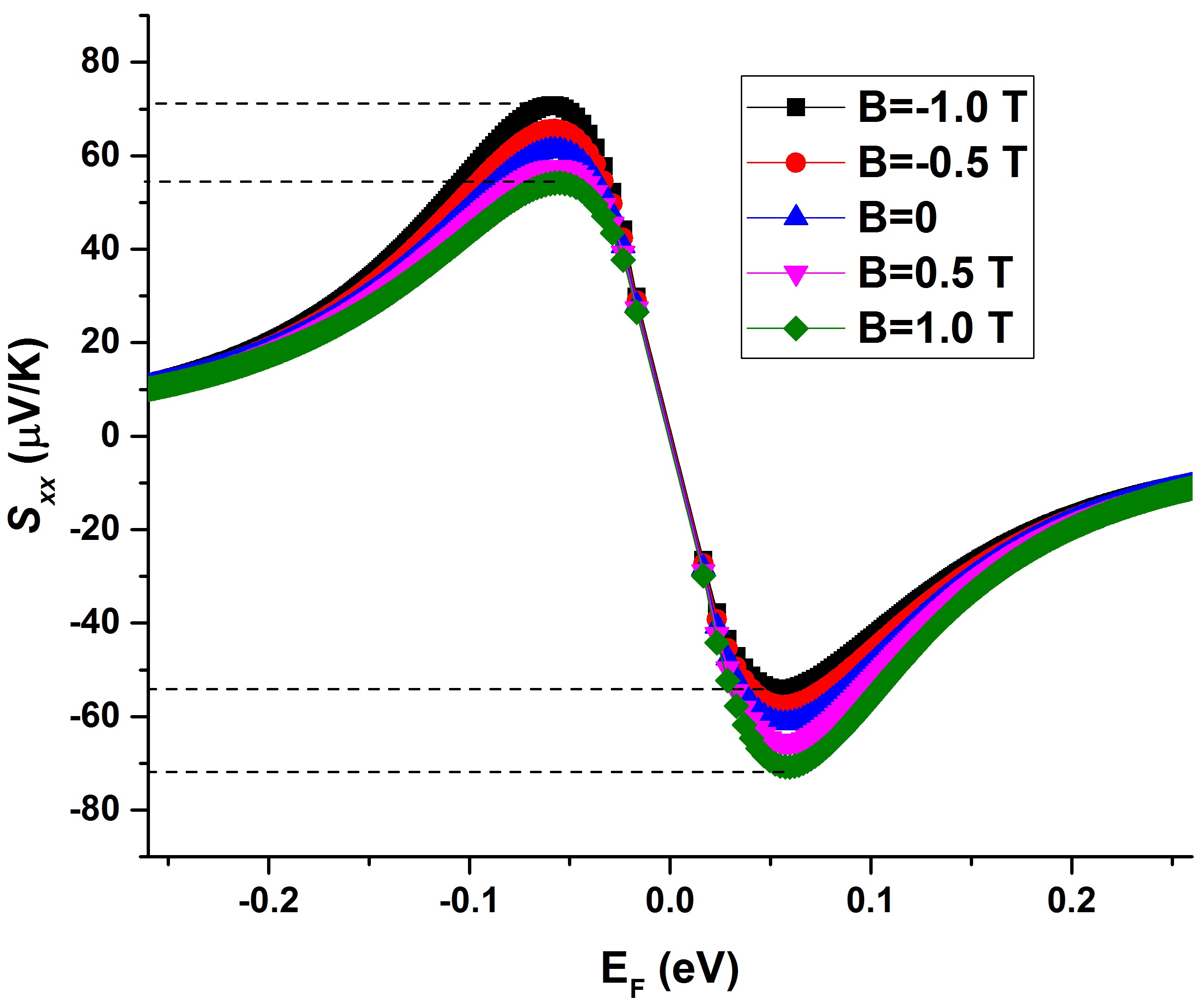}
\label{Fig10}
\caption{Variation of Seebeck coefficient $S_{xx}$ with magnetic field $B$ from -1 T to 1 T at temperature $T$=300 K and $n_{imp}=n_{vac}$=2.5$\times$ 10$^{11}$ cm$^{-2}$. } 
\end{figure}

\begin{figure} 
\includegraphics[width=9cm, height=7cm]{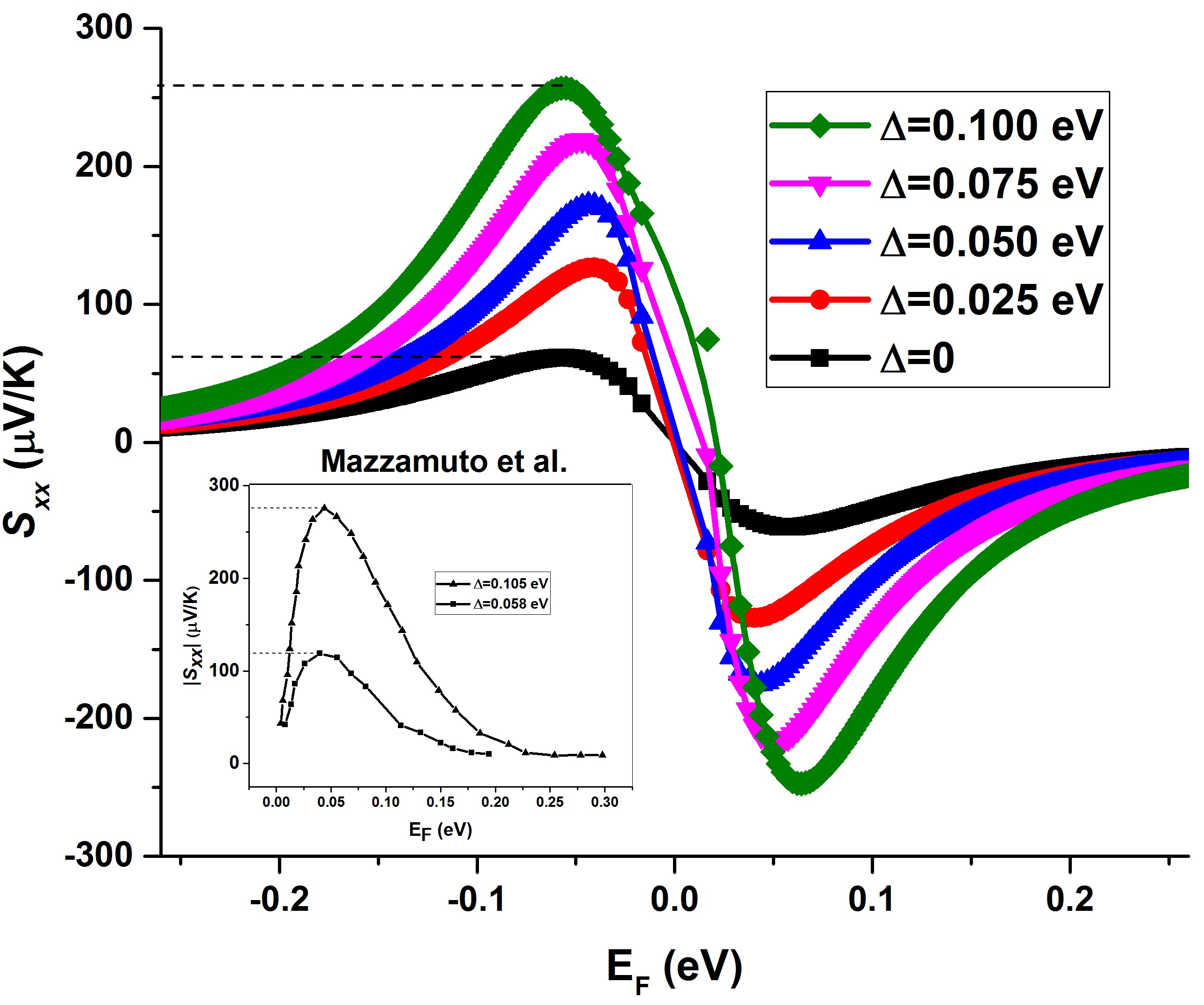}
\label{Fig11}
\caption{Variation of Seebeck coefficient $S_{xx}$ with Fermi energy $E_F$ for band gap $\bigtriangleup$ varying from 0 to 100 meV at temperature $T$=300 K, magnetic field $B$=0 and charged impurity and vacancy concentrations $n_{imp}=n_{vac}$=2.5$\times$ 10$^{11}$ cm$^{-2}$. The inset shows the results from the non-equilibrium Green's function model of Mazzauto et al. \cite{2011_Mazzamuto} for band gap $\bigtriangleup$ value of 56 meV and 105 meV. } 
\end{figure}

\begin{figure} 
\includegraphics[width=9cm, height=7cm]{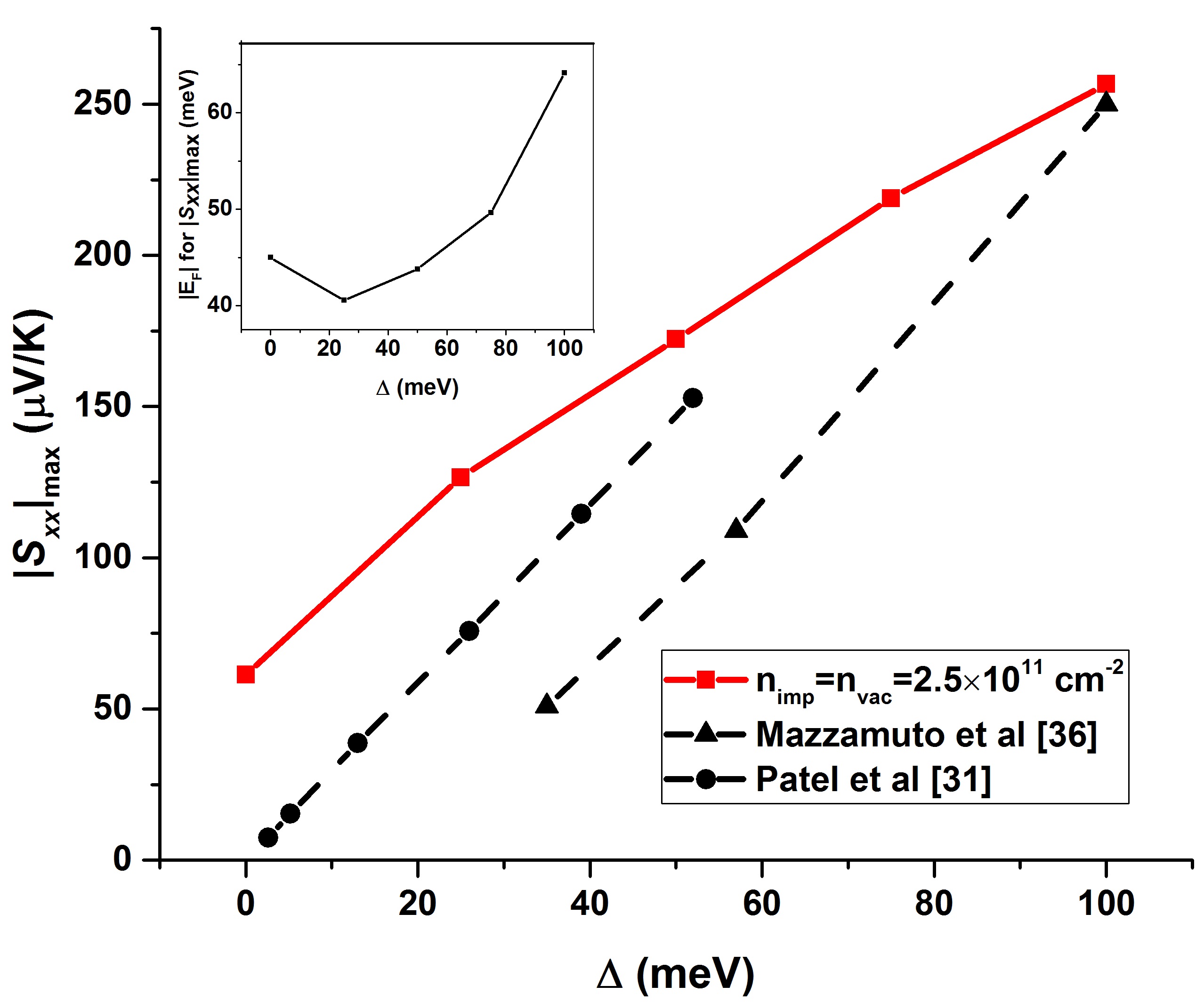}
\label{Fig12}
\caption{Variation of peak value of Seebeck coefficient $S_{xx}$ with band gap $\bigtriangleup$ ranging from 0 to 100 meV at temperature $T$=300 K, magnetic field $B$=0 and impurity concentrations $n_{imp}$=$n_{vac}$=2.5$\times$ 10$^{11}$ cm$^{-2}$. The dotted lines show Green's function model results for graphene nanoribbons \cite{2011_Mazzamuto} and analytical model results for periodic single layer graphene \cite{Patel2012}. The inset shows the Fermi energy $E_F$ corresponding to the peak value of Seebeck coefficient $S_{xx}$. } 
\end{figure}

\renewcommand{\thefigure}{A1}

\begin{figure} [htp]
\includegraphics[width=8.5cm, height=6cm]{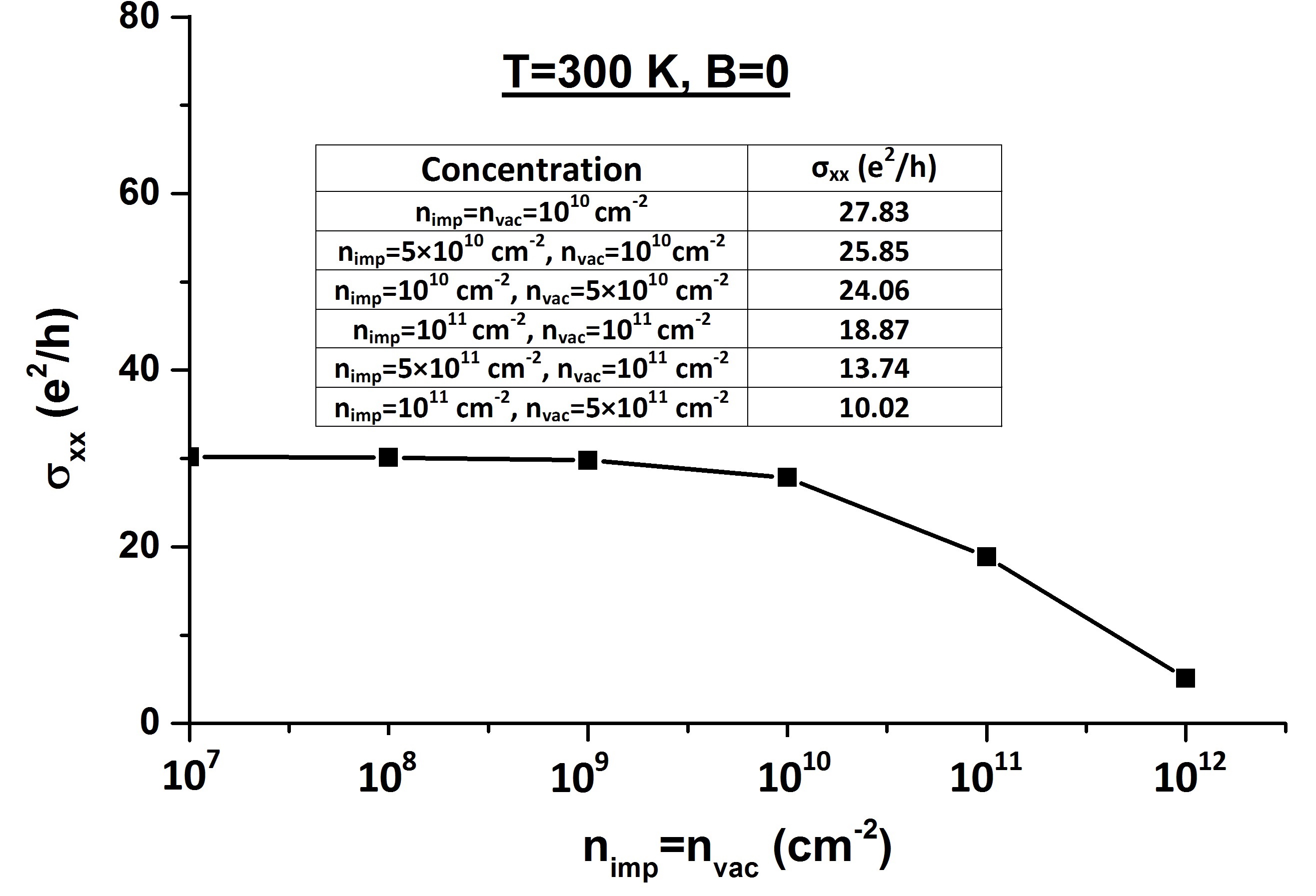}
\label{FigA1}
\caption{ Room temperature (300 K) longitudinal conductivity $\sigma_{xx}$ as a function of equal charged impurity and vacancy concentrations. } 
\end{figure}

\clearpage
 
\section*{Tables}
\begin{widetext}
\begin{table}[h!]
     \begin{center}
\caption{Summary of experiments on thermomagnetic effects in graphene.}
     \begin{tabular}{ |c| c|c|c|c|c|c|c|c|  }
     \hline

Method & \multicolumn{4}{c|}{Mechanical Exfoliation} &  \multicolumn{4}{c|}{Chemical Vapor Deposition (CVD)} \\ \hline
Layer & Single & Single & Single & Single & Single & Few & Few & Few \\ \hline
Temperature (K) & 10 - 300 & 1.5 - 300 & 20 - 280 & 20 \& 150 & 4 - 300 & 300 - 575 & 77 - 300 & 2 - 150 \\ \hline 
Gate Voltage (V) & -40 - +40 & -60 - +60 & -20 - +40 & -3 - +20 & -50 - +50 & - & - & -20 - +40 \\ \hline 
Magnetic  Field (T) & 8.8 & 1-8 & 5-14 & 1-8 & - & - & - & 13
\\ \hline
Carrier mobility  (cm$^2$/(V.s)) & 1000-7000 & 3000-12900 & - & 4560 - 17000 & 1500 - 13000 & - & - & 650 \\ \hline
$S_{xx}$  ($\mu$V/K) & 5-100 & 5-50 & 5-90 & 65-125 & 15-75 & 20-700 & 5-12 & 15-30 \\ \hline
$N$ ($\mu$VK$^{-1}$T$^{-1}$) & 30 & 5-45 & 20-40 & 50-250 & - & - & - & 30 \\ \hline
Reference & \cite{Zuev2009} & \cite{Wei2009} &  \cite{Checkelsky2009} & \cite{Liu2012} & \cite{Wang2011} & \cite{2011_Xiao} & \cite{2013_Babichev} & \cite{Nam2014}\\ \hline
\end{tabular}

\end{center}
\end{table}
\end{widetext}

\end{document}